\documentclass{WileyMSP-template}
\usepackage{graphicx}
\usepackage{dcolumn}
\usepackage{bm}
\usepackage{subfigure}
\usepackage{soul}
\usepackage{xcolor, soul}
\sethlcolor{green}
\sethlcolor{red}

\usepackage{subfigure}
\usepackage{graphicx,epsfig,epstopdf}
\usepackage{amsmath}
\usepackage{color}
\usepackage{gensymb}
\usepackage{mathtools}

\begin{document}


	\title{Conditional Generative Adversarial Networks for Inverse Design of Multi-functional Microwave Metasurfaces}
	
	\maketitle
	
	
	\author{Mehdi Kiani} and\author{Jalal Kiani*}\\
	
	\begin{affiliations}
		M. Kiani\\
		School of Electrical Engineering, Iran University of Science and Technology, Tehran, Iran\\
		
		J. Kiani, Ph.D.\\
		FedEx, Memphis, TN 38125, USA\\
		Email Address: jalalkianii@gmail.com
	\end{affiliations}
	
	

	\begin{abstract}
		\justify
		Electromagnetic (EM) metasurfaces can present a versatile platform for realization of multiple diverse EM functionalities with incident wave frequency, polarization, propagation direction, or power intensity through appropriate choice of unit cells structures. However, the inverse design of multi-functional metasurfaces relies on massive full-wave EM numerical simulations to obtain an optimized solution. This article proposes a step-by-step procedure based on conditional Generative Adversarial Networks (cGANs) integrated with Gramian  Angular Fields (GAFs) to reduce the computational time required for the EM simulations in the inverse design of multi-functional microwave metasurfaces. The proposed procedure initially implements GAFs to encode the desired multi-objective scattering parameters to images and then passes them through the cGAN model to map them to three-layer metasurfaces. The present study uses a robust dataset with a wide range of values and designs, including 54,000 metasurface structures and corresponding scattering parameters to train and validate the cGAN model. This article also presents a case study example using a multi-functional metasurface with three independent functionalities and full-space coverage to justify the performance of the proposed procedure in the inverse design of multi-functional microwave metasurfaces. The results demonstrate that the performance of the proposed model is promising, and the proposed procedure can be used to efficiently and accurately design multi-functional metasurfaces.
		
	\end{abstract}
	
	\begin{figure*}[t]
		\centering
		\includegraphics[height=3.3 in]{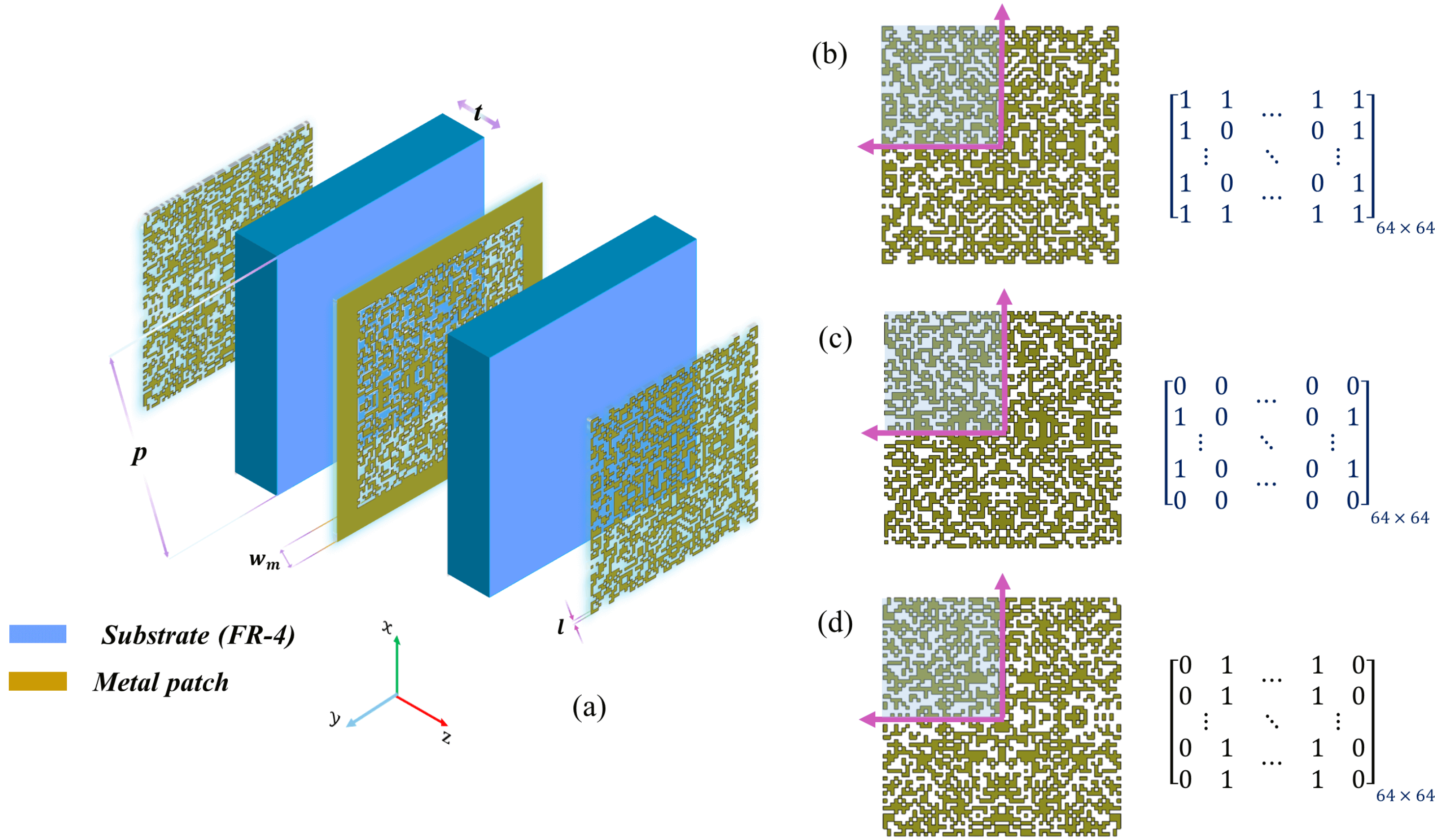}
		\caption{\label{fig:epsart} {(a) The perspective view of the three-layer unit cells contains 64$ \times $64 configurations of metal and void square blocks. (b) Top layer, (c) middle layer, and (d) bottom layer.  
		}}
	\end{figure*}
	\section{Introduction}
	\justify
	Electromagnetic (EM) metasurfaces are two-dimensional counterparts of EM metamaterials that received substantial attention for their great flexibility in EM waves control \cite{chen2016review, li2018metasurfaces}. They represent a thin-layer periodic or quasi-periodic array of metallic and/or dielectric elements, called unit cells, tailored to control the EM waves in unconventional ways. The realization of these structures has brought on a multitude of applications in achieving mantle cloaks \cite{chen2011mantle,sounas2015unidirectional}, polarizers \cite{zhao2011manipulating, kiani2020spatial}, wave-front manipulators  \cite{monticone2013full,yu2011light}, holograms   \cite{zheng2015metasurface,li2017electromagnetic}, analogue computers \cite{pors2015analog, babaee2021parallel}, absorbers \cite{liu2017experimental,kiani2020self}, flat lenses \cite{aieta2012aberration,khorasaninejad2017metalenses}, etc. 
	
	A critical issue concerning metasurfaces is designing appropriate unit cells to reach the required scattering coefficients \cite{saenz2009coupling}, equivalent refractive indexes \cite{quevedo2015ultrawideband}, or surface impedances \cite{sievenpiper1999high} for performing the applications. The traditional design of unit cells requires a deep understanding of the EM fields distribution in the unit cells metallic patterns to make an initial design. Additionally, the design process necessitates establishing tremendous numerical EM simulations and solving the expensive wave’s equations iteratively to reach the optimized EM responses. On the other hand, the design of metasurfaces with multiple functionalities needs more complex unit cells. Therefore, the design process of multi-functional unit cells is accompanied by highly time-consuming and costly optimization of a broad range of unit cell's geometric parameters.
	
	Over the past decade, due to the remarkable developments in data science and big data computing, the applications of machine learning methods across many fields, including computer vision \cite{krizhevsky2012imagenet, voulodimos2018deep}, natural language processing \cite{nadkarni2011natural}, speech recognition \cite{deng2013new}, quantum computing \cite{biamonte2017quantum} reinforcement learning \cite{krizhevsky2012advances}, and in different fields of engineering \cite{Kiani2019, campbell2020explosion, raccuglia2016machine, Kiani2020} are continuously growing. Recently, several research studies were published in the field of optical metamaterials and metasurfaces to implement machine learning tools to automatically design and optimize chiral metamaterials \cite{Ma2018}, resonant metasurfaces \cite{zhou2021inverse}, multi-parametric metasurfaces \cite{yeung2021global}, metasurface imagers \cite{Li2019}, all-dielectric metasurfaces \cite{nadell2019deep}, intelligent coding metasurfaces \cite{liu2021intelligent}, multi-functional metasurfaces \cite{an2021multifunctional}, etc. In these studies, the machine learning algorithms helped reduce the computational time required for the numerical simulations of classical metamaterials and metasurfaces. Inspired by the recent rapid advancements of machine learning tools in optics, several research studies elaborately explored machine learning-based inverse design of microwave metasurfaces to overcome the sophistic challenges of metasurfaces design \cite{Qiu2019, Zhang2019, hodge2019rf, naseri2021generative, mohammadjafari2021designing, shi2020metasurface, zhu2021phase}.
	
	Qiu et al. \cite{Qiu2019} introduced a deep learning-based approach for microwave metasurfaces inverse design. Their proposed model automatically generates a one-layer unit cell given a desired reflection amplitude response as the input. In addition, owing to the recent advancements of the reprogrammable digital coding metasurfaces in the engineering of EM waves \cite{cui2014coding, momeni2018information, liu2017concepts, rouhi2021multi}, a deep learning technique for image processing was proposed for inverse design of anisotropic digital coding metasurfaces \cite{Zhang2019}. This deep convolutional neural network could automate designs from the desired reflection phase to the objective unit cell structure. However, these metasurfaces inverse design models can only realize the target design in proportion to the amplitude or phase of reflection. Therefore, they cannot generate unit cells for different plausible applications requiring simultaneous control over both amplitude and phase. 
	
	Generative models were used as a solution to this shortcoming in the metasurfaces inverse design \cite{hodge2019rf, naseri2021generative,  mohammadjafari2021designing}. Hodge et al. \cite{hodge2019rf} trained a deep convolutional generative adversarial network on a dataset of known unit cells to generate new unit cells tailored to the design objectives. Similarly, a generative machine learning model based on a variational autoencoder for designing multi-layer metasurfaces in transverse-electric (TE) and transverse-magnetic (TM) polarizations was developed \cite{naseri2021generative}. However, these designs were trained using variations in geometric parameters of a limited number of known unit cells such as Jerusalem cross, rectangular and elliptical patches, and ring resonators that restrict the access to various EM functionalities by the designed metasurfaces. Moreover, these models only design metasurfaces in one-half EM space, leaving half of the space entirely unused, or with a single EM function that is unattractive for future smart radio environments and adaptive wireless networks.	
	
	The ever-increasing demand for system integration and device miniaturization, specifically in wireless communications, led to growing interest in multi-functional metasurfaces. One way to address this need is to employ semiconductor lumped elements into the structures to reach reconfigurability \cite{tang2020wireless, li2017nonlinear,momeni2022switchable}. Reconfigurable metasurfaces can show various EM functionalities depending on the controlling signals. However, they usually need DC bias networks which dramatically increases the complexity of the systems. On the other hand, designing multi-functional passive metasurfaces that cover the whole space is time-consuming and computationally expensive. Using machine learning-based approaches for the inverse design of multi-functional passive metasurfaces can significantly reduce the computation and optimization time of the design process. In this regard, an approach based on conditional Generative Adversarial Networks (cGANs) is developed for the inverse design of dependent and independent multi-functional passive microwave metasurfaces for full-space coverage. The proposed methodology encodes the scattering parameters to images using Gramian Angular Fields (GAFs), allowing the cGAN model as a computer vision tool to be used for designing multi-functional metasurfaces with full-space coverage. The training and validation datasets are unit cells structures in the form of 64$ \times $64$ \times $3 matrices of metallic and void square blocks with various metallic square block presence probabilities (MSBPPs) and metal and void square block arrangements (MVSBAs) and their corresponding scattering parameters with a broad range of values. For validation of the model, a passive microwave metasurface is presented with three independent functionalities for EM wave manipulation in the full space at a fixed frequency.
\begin{figure*}[t]
		\centering
		\includegraphics[height=3.5in]{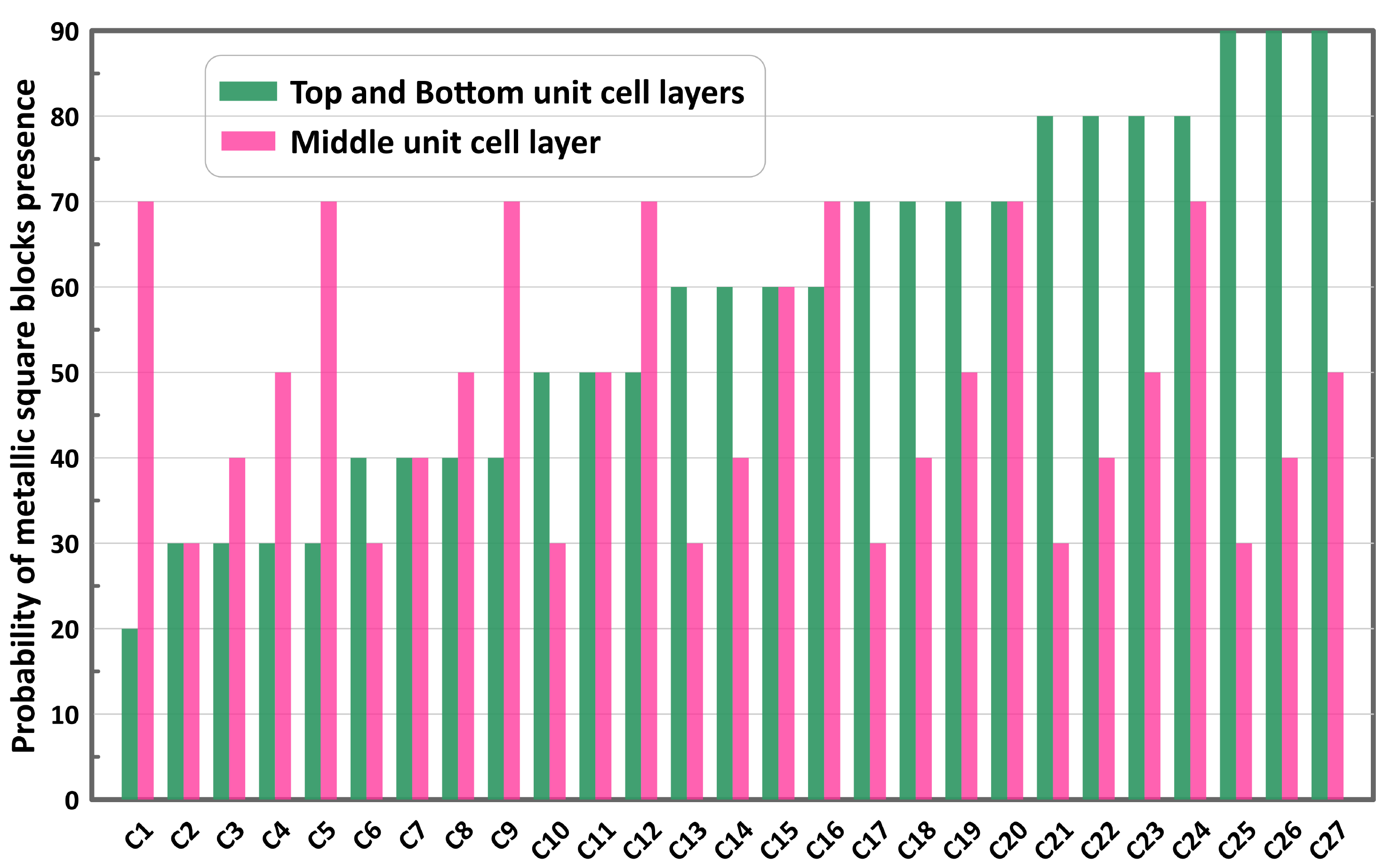}
		\caption{\label{fig:epsart} {The MSBPP distribution in the three layers of the studied unit cells in the different cases.}}
\end{figure*}
\section{Study Dataset}
	The study dataset is a collection of several components: metasurfaces characteristics, EM features, and the metasurfaces responses exposed to EM waves. The dataset is built using results from analyses of about 54,000 configurations of microwave metasurfaces excited by EM waves across an 8-12 GHz frequency range. Each metasurface unit cell in the study dataset comprises a 64$ \times $64$ \times $3 configuration of metal and void square blocks. The diversified MSBPPs and MVSBAs in these unit cells of the dataset result in a reliable and robust training dataset. In this study, CST Microwave Studio EM simulator \cite{CST} is used to calculate the scattering parameters of the metasurfaces as the EM responses.

	The following parts provide additional information about the overall specifications of the proposed metasurfaces, incident EM waves, and the data collection process.
	\subsection{Metasurface Model}
		This study employs three-layer metasurfaces to manipulate EM waves in the full-space. As shown in \textbf{Figure 1}, the three-layer metasurfaces unit cells are accomplished by means of three configurations of 64$ \times $64 square blocks (or a 64$ \times $64$ \times $3 configuration) with side length $l = 0.125$ $ mm $, situated on FR-4 dielectric substrates with thickness $ t = 1.5 $ $ mm $, period $ p = 10 $ $ mm $, dielectric constant $\epsilon_{r} = 4.3$ $F/m$, and loss tangent $tan\delta = 0.025$. In order to create a slot-based coat, a thin metallic square ring with a width $ w_m = 2$ $ mm $ is patched in the middle layer so that all around the middle layer is covered by the metal. The 64$ \times $64$ \times $3 configurations are composed of two types of metallic blocks and void blocks, which are distributed randomly on the unit cells.
	\subsection{Electromagnetic Excitation}
		In general, for the design of metasurfaces with single functionality, one parameter, the reflection coefficient or transmission coefficient of the metasurface in just one polarization, is usually required to be considered. However, to design metasurfaces with several functionalities, examining the reflection and transmission coefficients in TE and TM polarizations is necessary.  In this study, TE- and TM-polarized EM plane-waves illuminate the three-layer unit cells in the $ -z $ and $ +z $ directions. The reflection and transmission coefficients of the structures in these polarizations and directions are computed by CST Microwave Studio. Periodic boundary conditions (PBCs) are activated in $ x $ and $ y $ directions and Floquet ports are defined along the $ -z $ and $ +z $ directions in order to construct a transversely-infinite array from the unit cells shown in \textbf{Figure 1}. The reflection and transmission coefficients of metasurfaces ($ S_{11} $, $ S_{21} $, and $ S_{22} $) are examined to achieve metasurfaces with desired EM functions using the machine learning model.
	\subsection{Data Collection}
		It is well acknowledged that the performance of the deep learning models significantly depends on the quality and quantity of the training samples. When the amount of training samples is small, the deep learning models are likely to overfit the training data resulting in low performance for unseen data. 54,000 data samples that sufficiently represent the metasurface inverse design problem are collected to prevent the overfitting of the proposed deep learning model. The dataset contains 27 different cases, each containing 2000 64$ \times $64$ \times $3 configurations of square blocks and their scattering parameters. Each configuration of square blocks differs from the other configurations in at least one of these two aspects: first, the MSBPPs in each layer of the three-layer unit cells, and, second, the MVSBAs in the different layers of the unit cells.
		As shown in \textbf{Figure 2}, MSBPPs for the top and bottom layers change from 20\% to 90\%, whereas for the middle layer, they change from 20\% to 70\%. For example, in case 1, MSBPPs in the top and bottom layers are 20\%, while it is 70\% in the middle layer. Such a robust training dataset with various MSBPPs and MVSBAs significantly improves the proposed deep neural network performance. To observe and compare the performance of different cases in generating EM responses, the phase and amplitude distributions of the reflection and transmission coefficients are calculated and shown in \textbf{Figure 3}. This figure shows that different cases with different designs produce diverse amplitude and phase distributions. For example, the reflection amplitude and phase for case 25 mainly change between 0 and -1 dB and between -50$ ^{\circ} $ to -100$ ^{\circ} $, respectively. For case 5, these values are between 0 and -1.5 dB and -40$ ^{\circ} $ to -180$ ^{\circ} $, respectively. In fact, the difference in the amplitude (and phase) distributions of different cases is due to the difference in the amount of MSBPPs.
\begin{figure*}[!t]
	\centering
		\subfigure[][]{%
		\label{fig:23}%
		\includegraphics[height=1.9in]{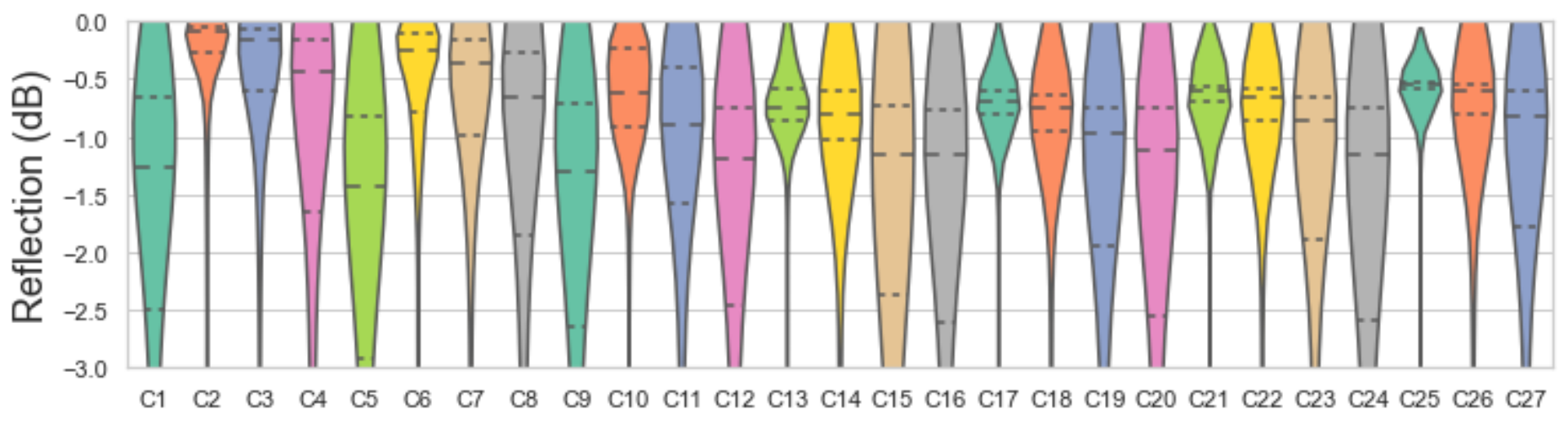}}%
	\qquad
		\subfigure[][]{%
			\label{fig:23}%
			\includegraphics[height=1.85in]{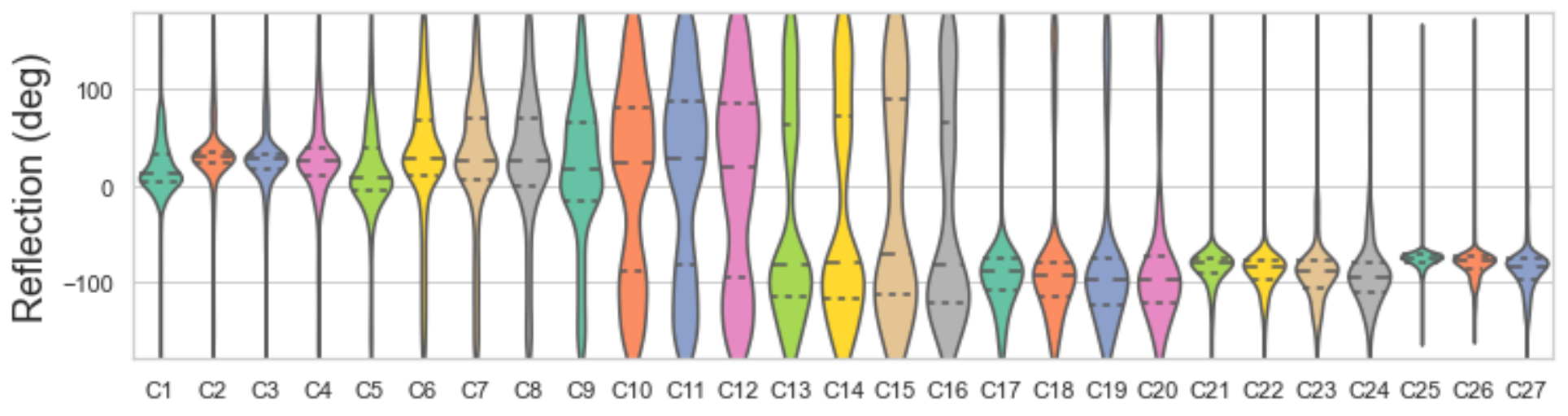}}%
		\qquad
		\subfigure[][]{%
			\label{fig:23}%
			\includegraphics[height=1.9in]{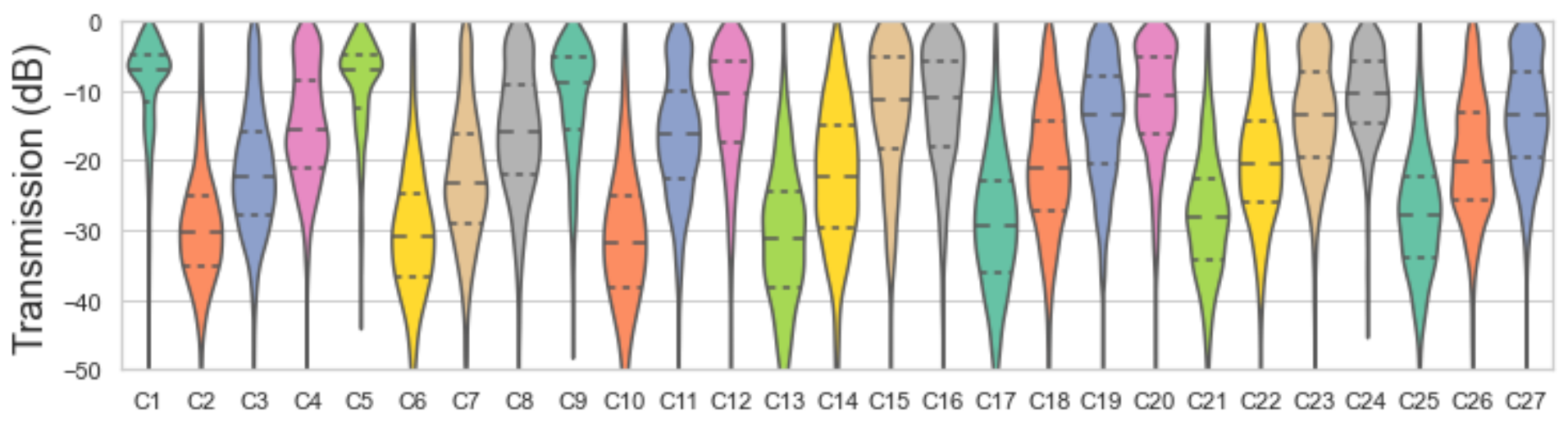}}%
		\qquad
		\subfigure[][]{%
			\label{fig:23}%
			\includegraphics[height=1.85in]{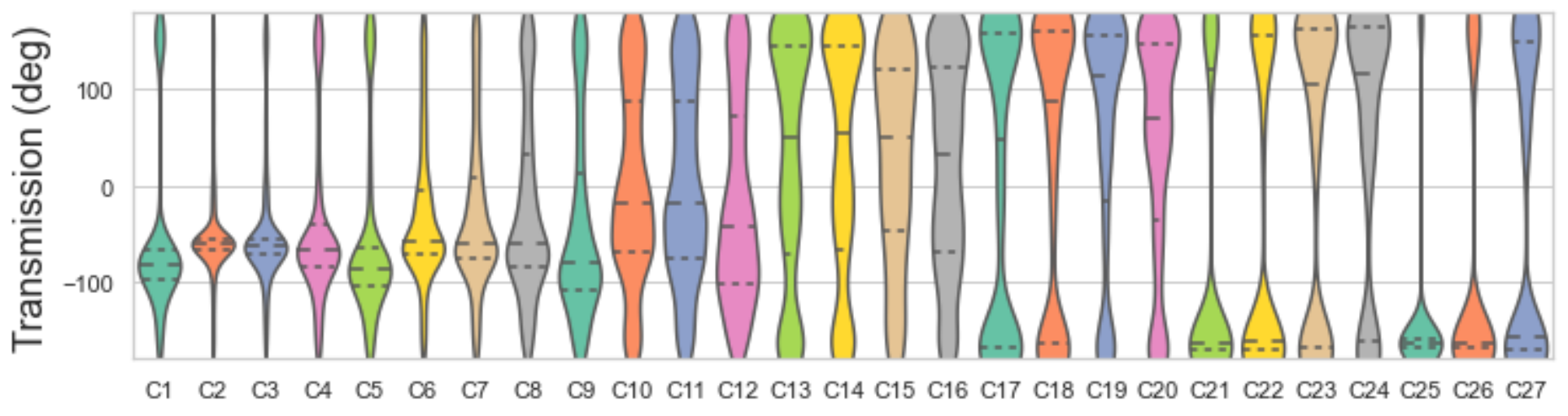}}%
		\caption{\label{fig:epsart} {(a) and (b) The amplitude and phase distributions of the reflection coefficient for the multiple cases. (c) and (d) The amplitude and phase distributions of the transmission coefficient for the multiple cases. }}
\end{figure*}	
\section{Methodology}
	Generative Adversarial Networks (GANs) are deep learning-based generative models in which two sub-models, namely the generator and the discriminator, contend against each other to learn a mapping from random noise vector $ z $ to output image $y$, $G: z$ $\rightarrow$ $y$ \cite{goodfellow2014generative}. The generator sub-model generates new conceivable images from the problem domain, while the discriminator model seeks to classify the generated images as either real or fake.
	
	An outstanding extension to the GANs is their utilization to generate associate output conditionally \cite{mirza2014conditional}. In conditional GANs (cGANs), the input is conditional on some auxiliary information such as an input image. This provides one of the most compelling applications of the GANs, namely image-to-image translation \cite{isola2017image}. Isola et al. \cite{isola2017image} introduced the concept of Pix2Pix as a type of cGAN for comprehensive image-to-image translation tasks. In the  Pix2Pix model, the generation of an output image is conditional on an input image; in fact, the cGAN model performs a mapping from input image $ x $ and random noise vector $ z $ to output image $y$, $G : \{x, z\} $ $\rightarrow$ $ y $. In such models,  the generator takes an image (source image) as the input and tries to generate a sample output (generated image or target image) that cannot be identified from the real images. Then, the discriminator, a deep convolutional neural network that performs conditional-image classification, takes both the generated image and the target image as the input and predicts whether the target image is a real or a fake conversation of the source image. Therefore, the cGAN objective with the generator G and the discriminator D can be interpreted as a min-max game, as follows:
	\begin{equation}
	\begin{aligned}
	\underset{G}{min}\:\underset{D}{max}\:V(D, G) \:= \: E_{y \sim p_{data}(y)}[log(D(y|x))] \;+ \:
	E_{z \sim p_{z}(z)}[log (1 - D(G(z|x))] 
	\end{aligned}
	\end{equation}
	in which, $ p_{z}(z) $ is prior input noise distribution, D(y$ | $x) is the probability that y  given x is a real data. It is supposed that $ z $ is a random noise vector; accordingly, $ D(G(z|x)) $ is the probability that the generated images comprise noise and input images attained from the real dataset \cite{isola2017image, zhu2017unpaired}.
	At the beginning of training, the discriminator can effortlessly apprehend the random fake output from the real target image. As a result, the  $ D(G(z|x)) $  value is low, resulting in a large $ log(1-D(G(z|x))) $. Thus, the generator model is updated to minimize the loss predicted by the discriminator for generated images marked as real images. As such, the generator is encouraged to generate more synthetic images similar to the target or original images.
	
	In this study, the collected scattering parameters are converted into images, which will be discussed in detail in the next section. On the other hand, the unit cells structures are in the forms of matrices with the shape of 64$ \times $64 or images with the shape of 64$ \times $64$ \times $3 from the metallic and void square blocks. Hence, a cGAN model (Pix2Pix model) is introduced for the inverse design of multi-functional metasurfaces. The scattering parameters images are used as the input (source images) of the cGAN model and unit cells images (target images) as the model's output. The cGAN model generator tries to generate the corresponding unit cells images of the desired EM responses or the scattering parameters, and the cGAN model discriminator attempts to distinguish the valid unit cells from the invalid unit cells. Finally, the generator model designs unit cells whose EM behaviour leads to expected EM responses after learning.
	\begin{figure*}[!t]
		\centering
		\includegraphics[height=7in]{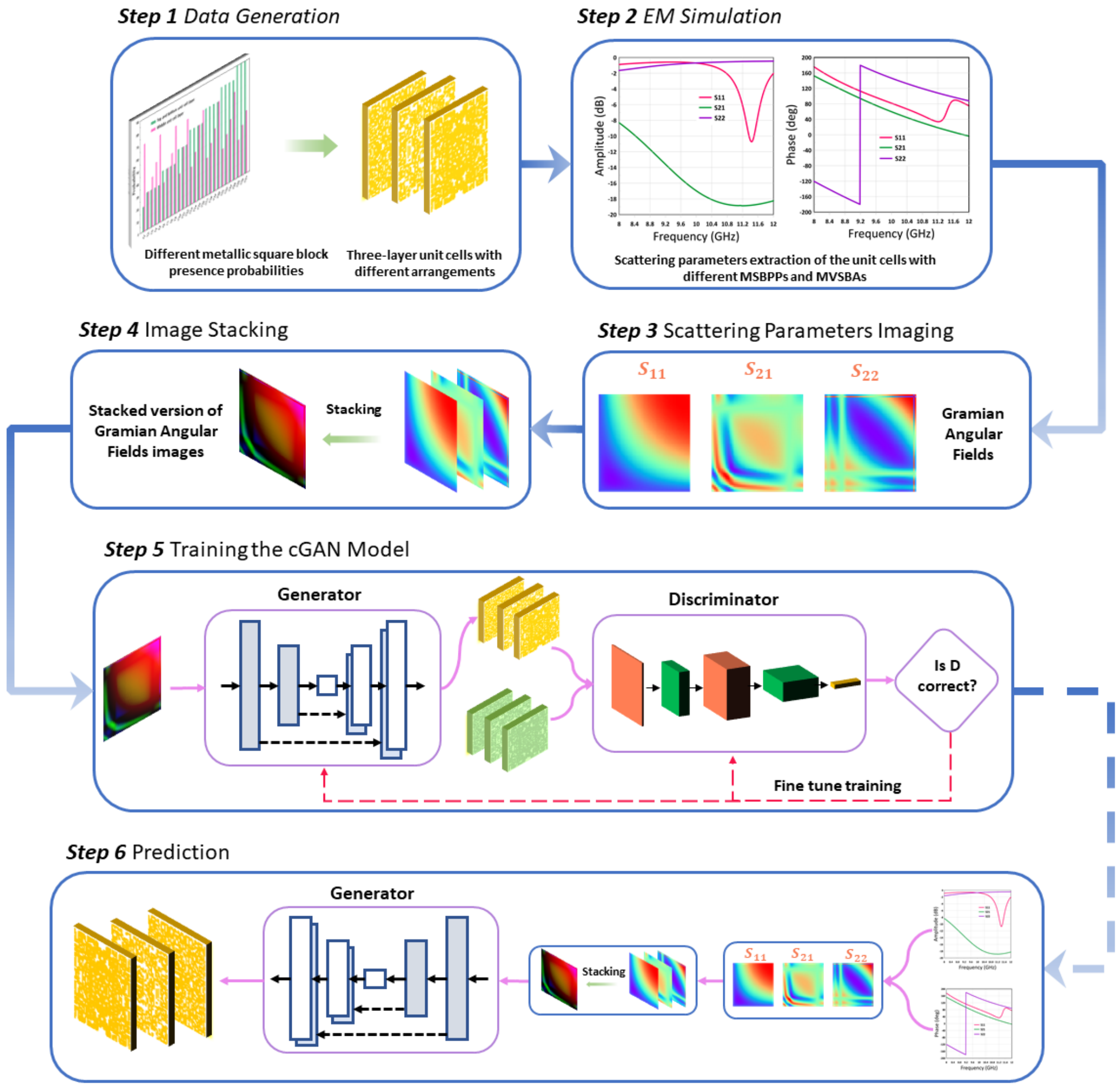}
		\caption{\label{fig:epsart} {The workflow of the cGAN-based methodology for the inverse design of multi-functional microwave metasurfaces.}}
	\end{figure*}
	\subsection{The Workflow of Applied Method}
	The workflow of the recommended methodology for inverse design of multi-functional metasurfaces is demonstrated in \textbf{Figure 4}. Achieving an accurate and reliable inverse design model based on this methodology comprises the following steps.
	
	\textbf{Step 1.} Data generation: generate data in terms of square blocks distribution in unit cells with various MVSBAs and MSBPPs. As discussed in Section II, it is significant to consider a wide range of unit cells with different metallic and void square blocks distributions.
	
	\textbf{Step 2.} EM simulation: extract the EM response of the produced unit cells. When choosing a set of unit cells, a key principle called diversity should be considered. Diversity means that the selected data samples must be distributed across the entire input space instead of focusing on a small local region \cite{wu2018pool}. In other words, the scattering parameters of the generated unit cells for TE and TM polarizations in $ +z $ and $ -z $ directions must cover a wide variety of values over the specified frequency range. Conversely, if the generated unit cells specify some limited EM responses, the machine learning method most likely learns less using the input/output samples within the local region. 
	
	\textbf{Step 3.} Scattering parameters imaging: encoding scattering parameters as GAFs images \cite{wang2015imaging}. Inspired by the great achievements of GANs in computer vision, there is a tendency to encode the scattering parameters as images so that machines can learn structures and EM responses visually. This article uses GAFs to encode the scattering parameters as images.  
	
	Gramian Angular Fields are images that depict a time series in a polar coordinate system, in which the location of each point $ P $ on the plane is indicated by its distance $ r $ from the origin and the angle $\phi$ relative to the positive x-axis. The scattering parameters calculated from step 2 can be considered as time series in the form of n observations $ S = \{ s_{1},s_{2} ,...,s_{n}\} $ in the frequency domain. On the other hand, the observations can be represented on the basis of Polar coordinates as follows:
	\begin{equation}
	\begin{aligned}
	r_{i} = \sqrt{Imaginary(s_{i})^{2} + Real(s_{i})^{2}},\:\:\: 0 \leq r \leq 1\\
	\phi_{i} = \tan^{-1}(\dfrac{Imaginary(s_{i})}{Real(s_{i})}), \:\:\: -180^{\circ} \leq \phi < 180^{\circ} 
	\end{aligned}
	\end{equation}
	
	Following the transformation of the scattering parameters into the Polar coordinate system, the GAFs calculate the inner product of the scattering parameters in the Polar system via the characterization of angular summation to detect the correlation across different frequency intervals. The GAFs of the scattering parameters are shown below.
	
	\begin{equation}
	\begin{aligned}
	GAF =	\langle S, \: \:S^{'}\rangle = S^{'} . \:\: S - \sqrt{I-S^{2}}^{'} . \: \: \sqrt{I-S^{2}}
	\end{aligned}
	\end{equation}\\
	in which I is the unit row vector [1, 1,..., 1].
	
	\textbf{Step 4.} Image stacking: stack scattering parameters images generated by GAFs, which are represented by three 2D 64 $\times$ 64 arrays, to produce three-layer images with the size of 64 $\times$ 64 $\times$ 3 pixels. The three-layer metasurfaces depending on the polarization and direction of the incident wave, show three independent applications at a fixed frequency. Hence, it is necessary to incorporate three scattering components to design the metasurfaces. To consider all three scattering components in the cGAN model, the three generated images of the scattering parameters are stacked and the resulted three-layer image is used as the input of the model.
	
	
	\textbf{Step 5.} Training the cGAN model: train both the generator model and the discriminator model as the main parts of the cGAN model simultaneously using the training data. In this step, the stacked versions (i.e., the 64$\times$64$\times$3 images) of the generated GAFs images, source images, are applied as the input of the generator model. The final expected outputs of the generator model are the three-layer metasurface unit cells, which are represented as three configurations of 64$\times$64 square blocks or 3D matrices. These generated outputs of the generator model and the original unit cells images, target images, are provided as the input of the cGAN discriminator model to distinguish the invalid outputs from the valid ones. The generator model is updated based on the calculated loss by the discriminator measuring the differences between the generated outputs and the expected or target outputs. The training continues until the cGAN objective in equation 1 is satisfied, or the discriminator could not identify the real unit cells images from the fake unit cells images. Of the 54,000 data points used for simulation in this study, 51,000 of them are implemented for the model's training and the rest (3,000 data points) are used for the validation of the model.
	
	\textbf{Step 6.} Prediction: predict the corresponding metasurface unit cells in regards to the desired EM responses.  When the cGAN model is trained using the scattering parameters and unit cells images, the
	discriminator cannot recognize the difference between the original unit cells structures and the generated unit cells given the input EM features. This means that the generator part can produce valid unit cells whose EM behaviours are the same as the input source EM features. Hence, the trained cGAN generator model can predict the proper unit cells configurations for the desired EM responses.
	
	\begin{figure*}[]
		\centering
		\includegraphics[height=2.8in]{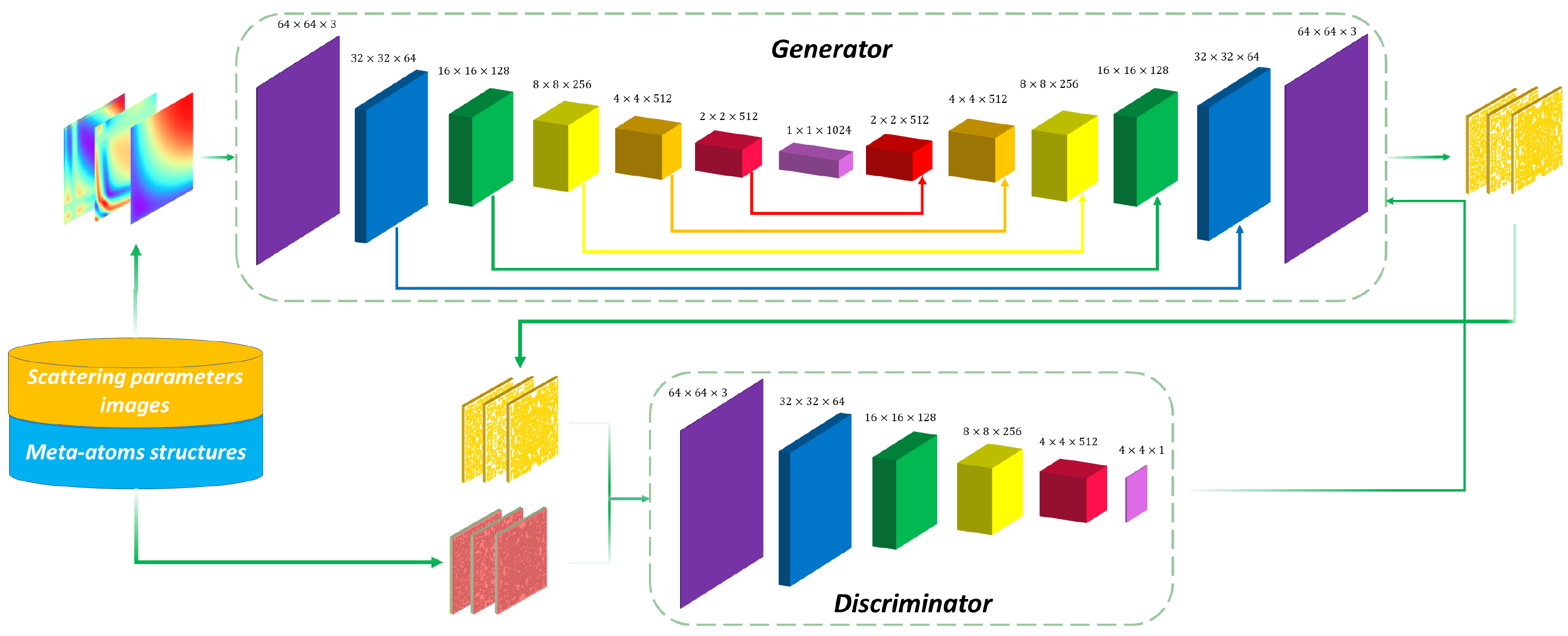}
		\caption{\label{fig:epsart} {The schematic diagram of the generator and discriminator layers in the proposed cGAN model.}}
	\end{figure*}
	\subsection{The Architecture of the Conditional Generative Adversarial Network}
	This part describes the technical aspects of the proposed cGAN model for the inverse design of multi-functional microwave metasurfaces. As shown in \textbf{Figure 5}, the cGAN model is comprised of two elements: the generator and the discriminator. The generator model uses the scattering parameters converted to stack images in the form of  64$ \times $64$ \times $3  as the input and generates the unit cells structures in terms of 64$ \times $64$ \times $3 images in which each channel represents one unit cell layer. In other words, the cGAN model passes the scattering parameter images through the generator network, which is an encoder-decoder model based on a U-Net architecture, to generate the unit cells structures. The symmetric U-Net architecture initially encodes the input images down to a bottleneck layer and then decodes the bottleneck representation to the size of the target images. General convolutional layers generate the encoder path and transposed 2D convolutional layers develop the decoder path \cite{ronneberger2015u}. Residual connections connect the convolutional layers with their corresponding transposed convolutional layers to prevent accuracy saturation in the deep convolutional neural network \cite{szegedy2017inception}. 
	
	According to \textbf{Figure 5}, the discriminator model is a deep convolutional neural network that performs conditional image classification. This study uses a PatchGAN paradigm as the discriminator that takes two types of input images, including the target images representing the original unit cells structures and the generated images. Then, it concatenates the images together, passes them through various hidden layers that decrease by a factor of 2, and predicts a patch output of predictions  \cite{isola2017image, zhu2017unpaired}. The output predictions of the proposed PatchGAN are 4$ \times $4 patches/portions of the unit cells images. The average of these output predictions gives an overall plausibility or classification score, which is used as feedback to the generator model to improve the quality of the generated unit cells images. 
	
	The implementation in this study uses the Keras deep learning framework for building the cGAN model, which takes the scattering parameters and generates metasurface structures with the size 64$\times$64$\times$3 pixels. Both the discriminator and generator models apply dropout regularization to mitigate overfitting and enhance the generalization of the deep convolutional neural network \cite{srivastava2014dropout}. An experimentation demonstrated that setting the epochs number and the batch size, respectively, to 150 and 1 leads to the best results as also recommended in \cite{isola2017image}. In addition, the cGAN model uses Adam optimizer with an initial learning rate of 3$\times 10^{-4} $, which has a high convergence speed and is quicker than gradient descent \cite{bengio2012practical, kingma2014adam}. Since there are 51,000 data points in the training dataset and the batch size is 1, each epoch involves 51,000 training steps. The generator is saved and evaluated every epoch or every 51,000 training steps, and the cGAN model will run for a total of 7,650,000 training steps (150 epochs $ \times $ 51,000 training steps).
	
	\begin{figure*}[]
		\centering
		\subfigure[][]{%
			\label{fig:23}%
			\includegraphics[height=1.6in]{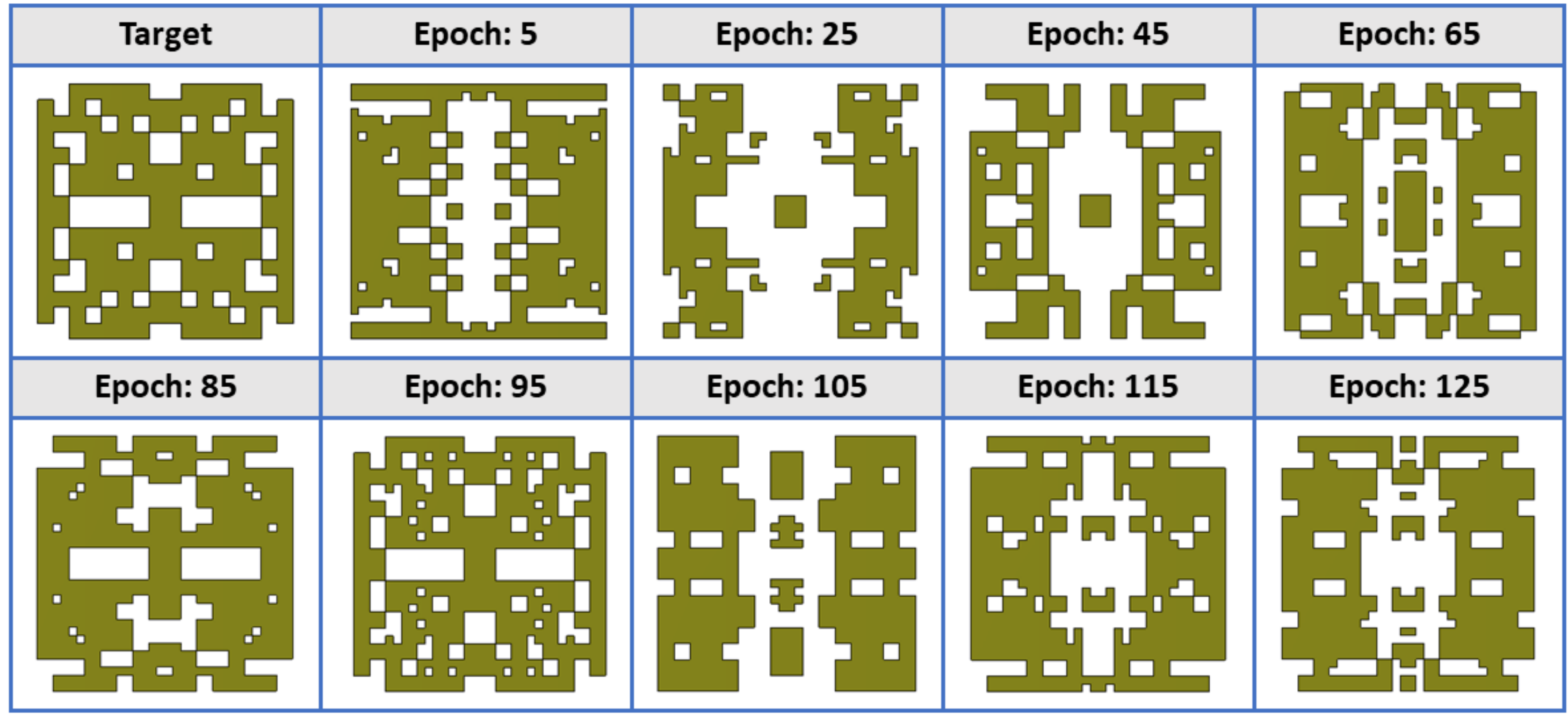}}%
		\qquad
		\subfigure[][]{%
			\label{fig:23}%
			\includegraphics[height=1.6in]{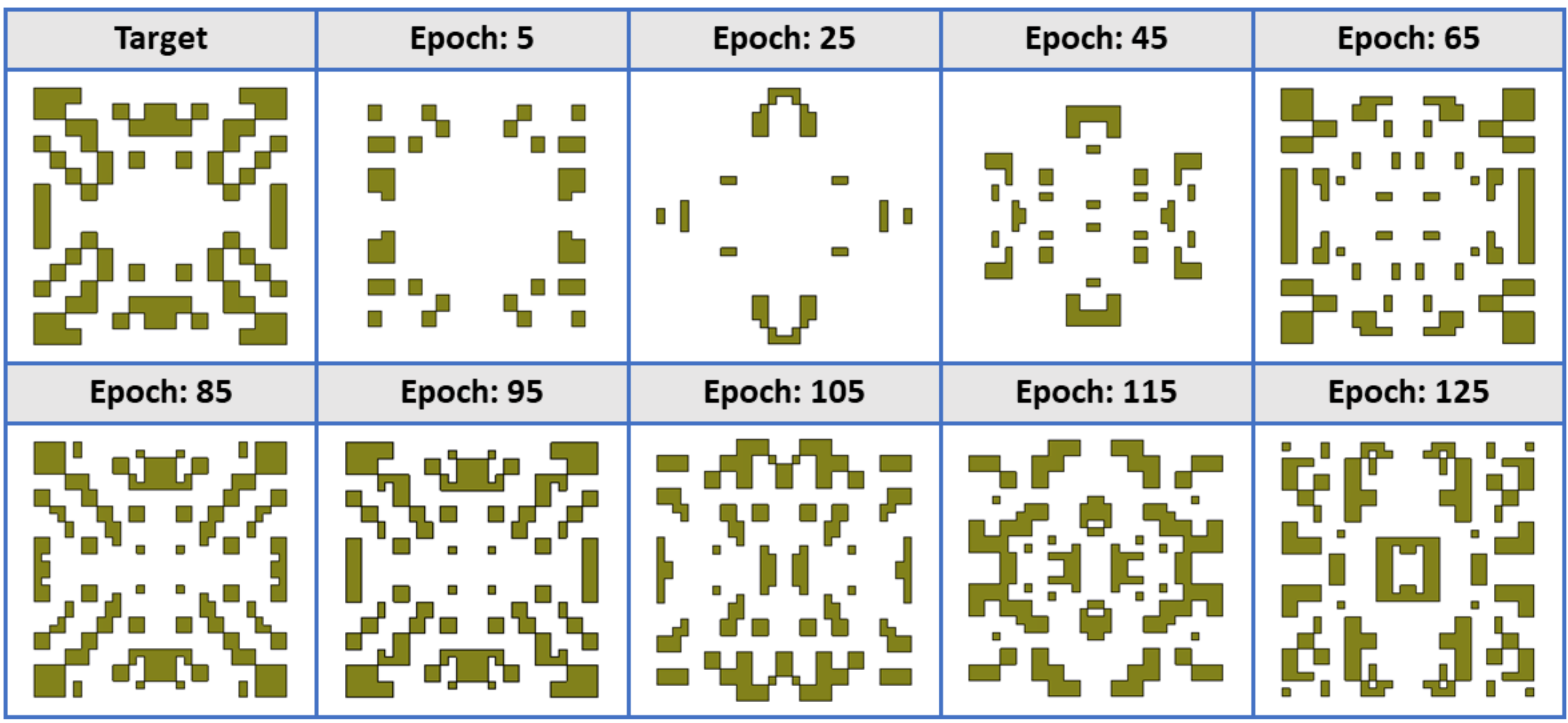}}%
		\qquad
		\subfigure[][]{%
			\label{fig:23}%
			\includegraphics[height=1.6in]{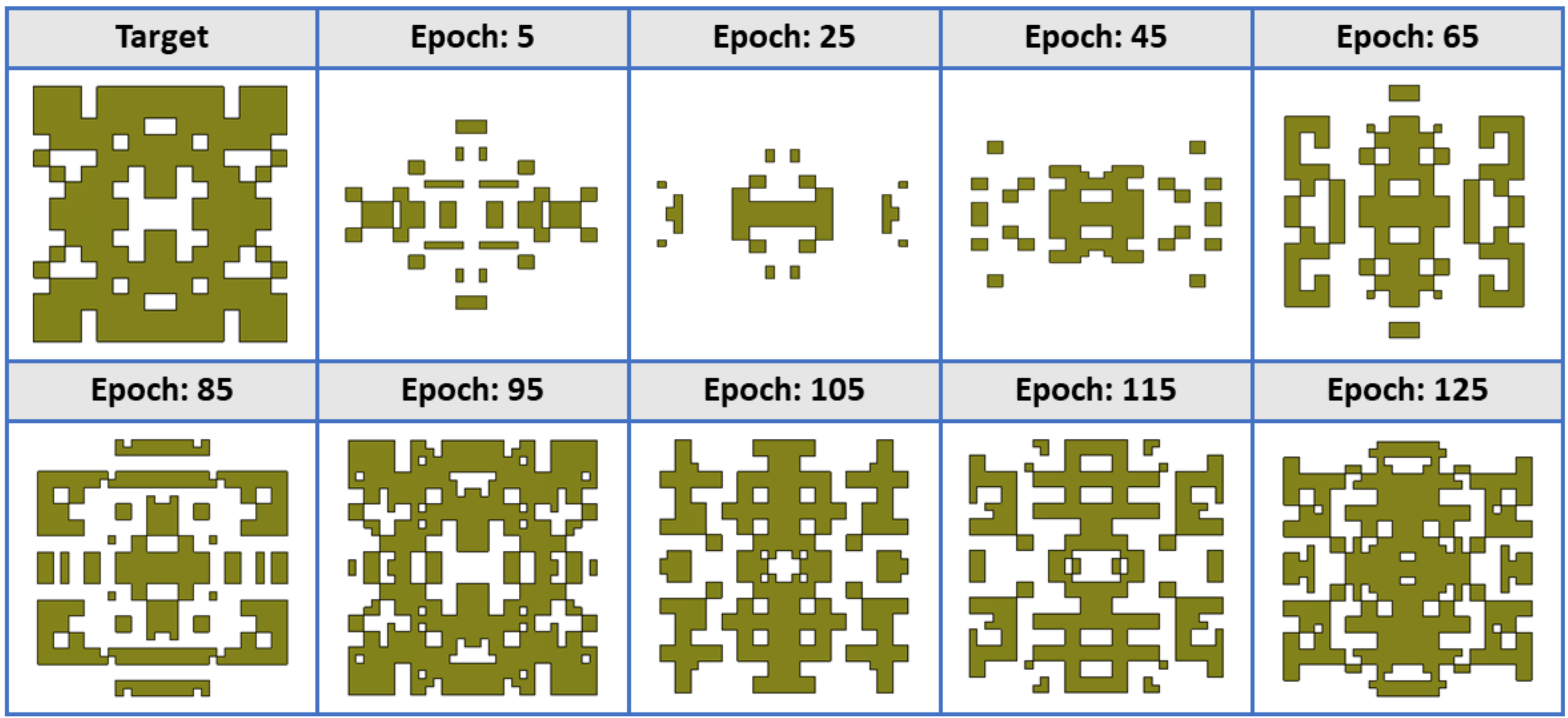}}%

		\caption{\label{fig:epsart} {The target and generated unit cell structures for a random example in the validation dataset at different epochs (a) layer 1, (b) layer 2, and (c) layer 3. }}
	\end{figure*}
	
	\begin{figure*}[]
		\centering
		\subfigure[][]{%
			\label{fig:23}%
			\includegraphics[height=1.6in]{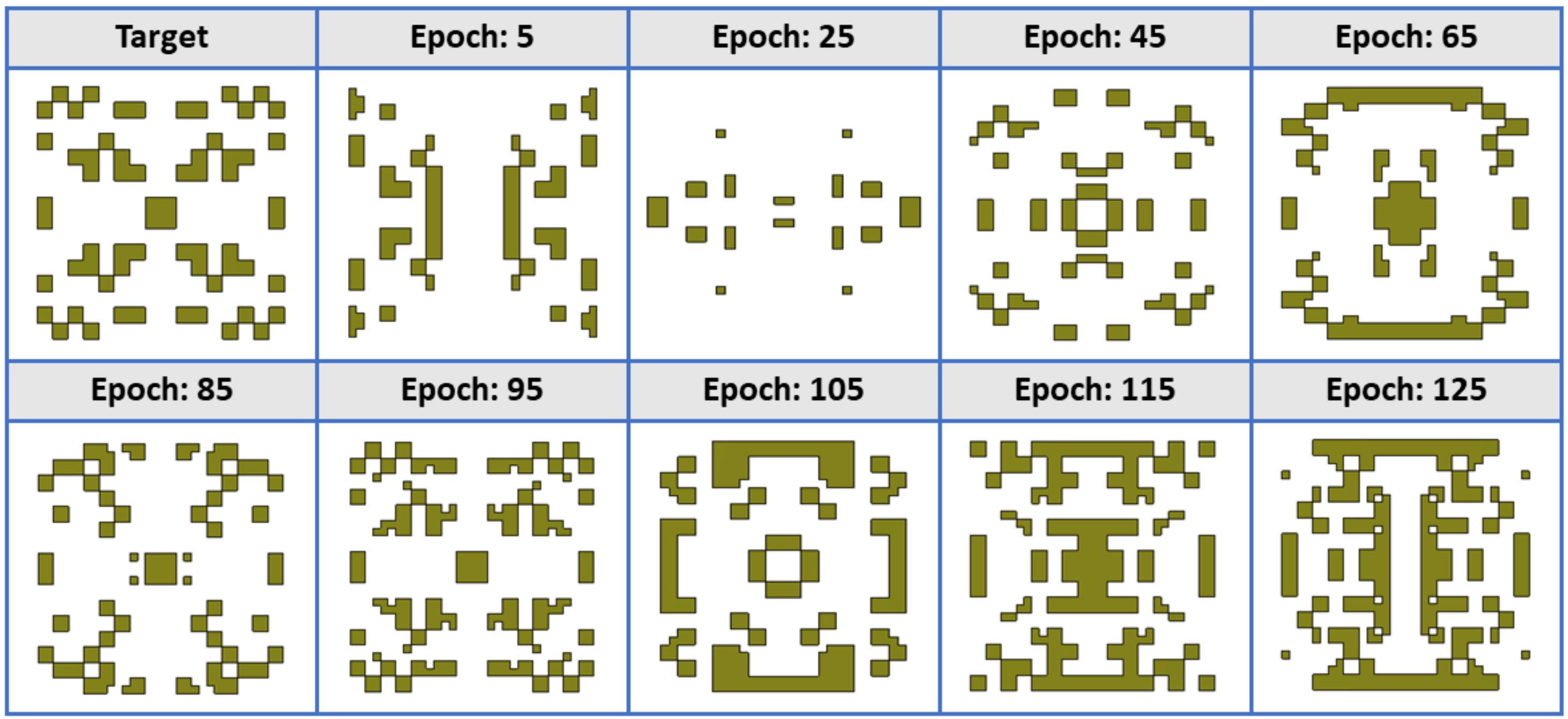}}%
		\qquad
		\subfigure[][]{%
			\label{fig:23}%
			\includegraphics[height=1.6in]{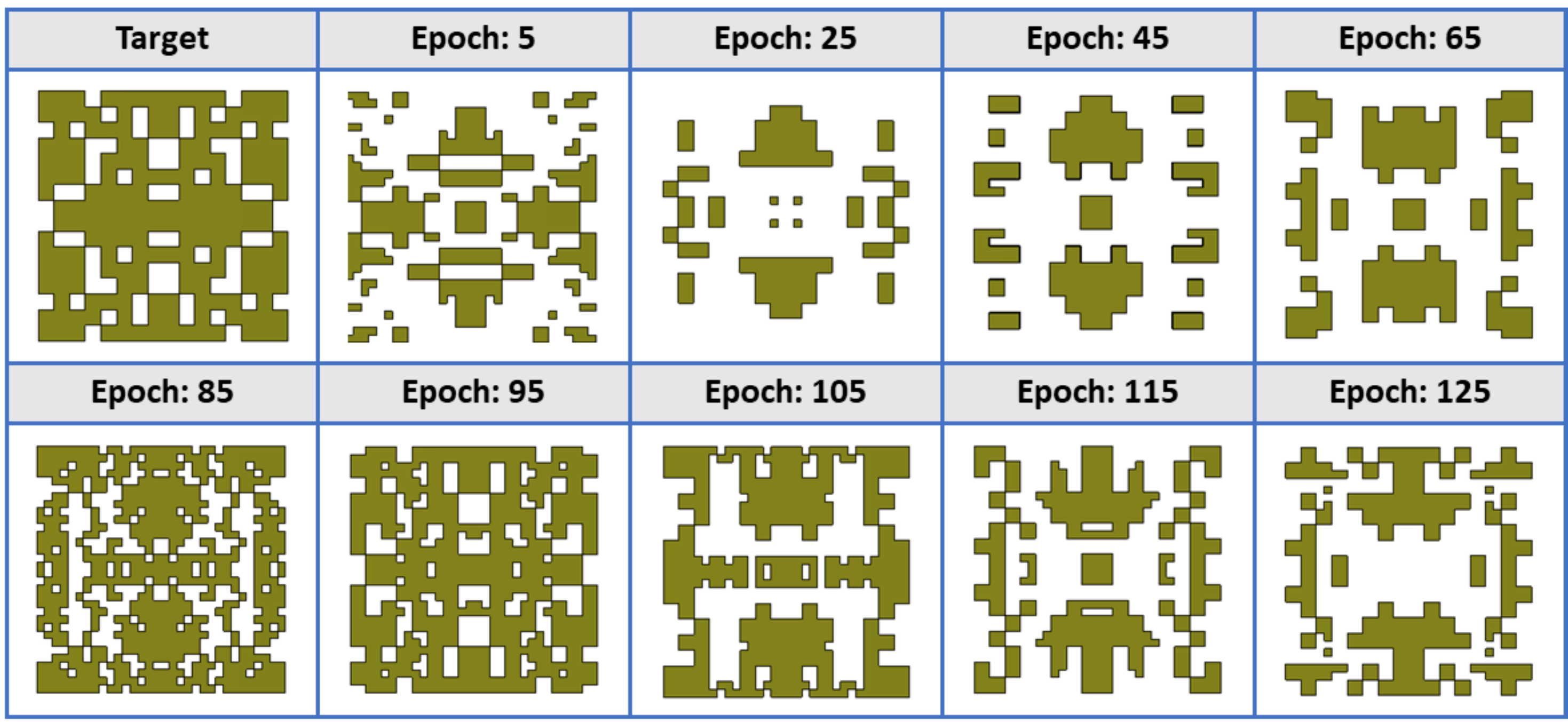}}%
		\qquad
		\subfigure[][]{%
			\label{fig:23}%
			\includegraphics[height=1.6in]{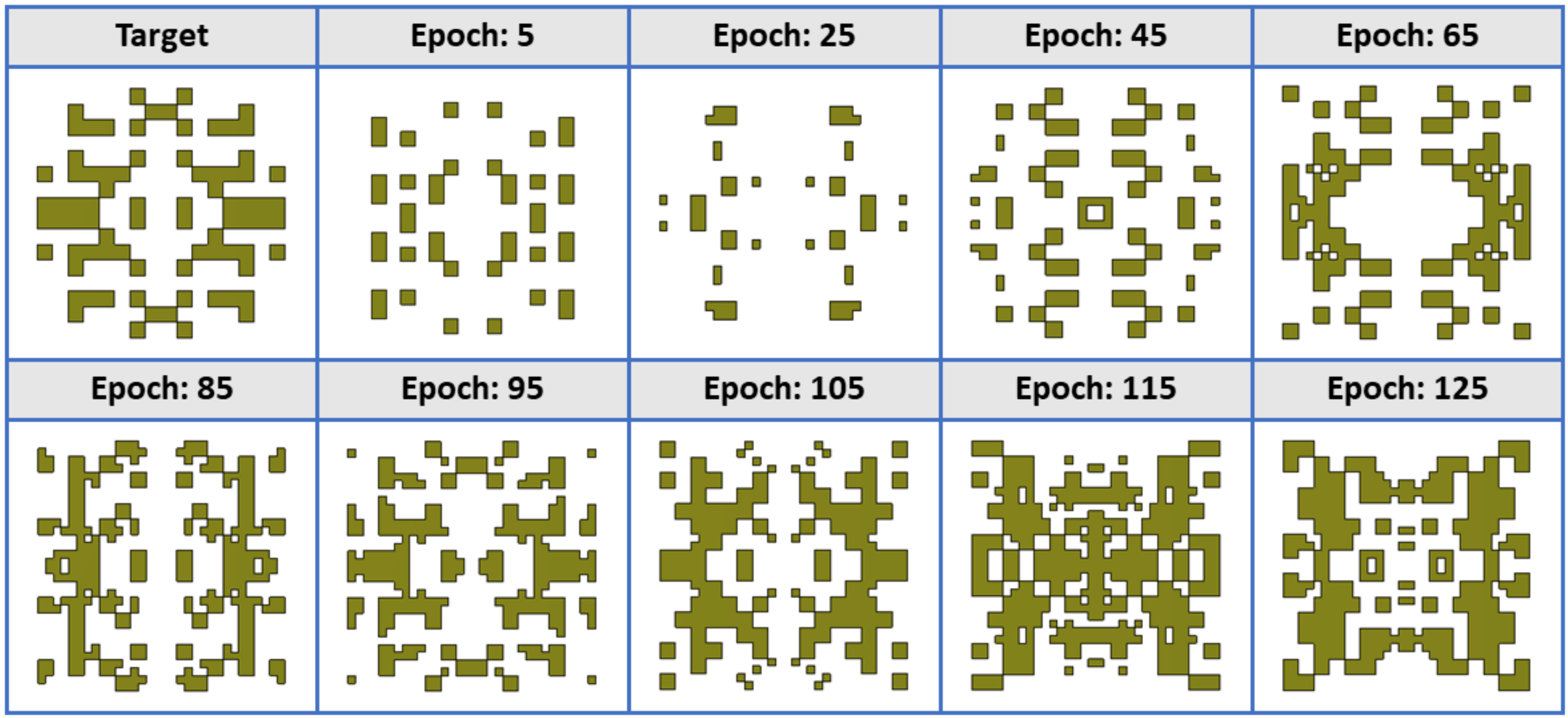}}%

		\caption{\label{fig:epsart} {The target and generated unit cell structures for a random example in the validation dataset at different epochs (a) layer 1, (b) layer 2, and (c) layer 3.}}
	\end{figure*}

	\section{Results and Discussion}
	\subsection{cGAN Model Performance Evaluation}
	
	In contrast to other deep learning models, evaluating the performance of the cGAN model and the quality of the generated images is a challenging open issue \cite{salimans2016improved}. Deep learning models usually use a loss function in the training process, while the generator in the cGAN models uses the loss calculated by the discriminator model. In these models, both the generator and discriminator are trained simultaneously by an adversarial process to reach a compromise. Therefore, the performance of cGAN model cannot be easily measured to decide when the training process should be stopped since it is not possible to use quantitative metrics to evaluate the performance of the cGAN models or assess the progress of the training. \cite{salimans2016improved}. In addition, metrics such as per-pixel mean-squared error cannot capture the very structure like structured losses because these metrics do not assess joint statistics of the generated images (or unit cells here). 
	
	To solve the mentioned issue, this study uses a "real versus fake" comparison technique through manual inspection of generated metasurfaces over different epochs to measure the quality of the generated unit cells and the efficiency of the cGAN model. Reviewing the generated images over different epochs helps to choose the best optimal model as the final model at the end of training. The "real versus fake" comparison technique involves using the cGAN model to generate a batch of synthetic unit cells structures given the input scattering parameters, then evaluating the quality and diversity of the generated unit cells structures compared to the original or target unit cells. For this purpose, this study compares different layers of the generated or fake unit cells versus the original unit cells over different epochs. To ensure that the cGAN model performs well on the unseen dataset, this study uses the validation dataset to measure the cGAN model's performance.  
	
	\textbf{Figures 6} and \textbf{7} show the generated unit cells structures over different epochs when the GAFs method is utilized for imaging the scattering parameters for two randomly selected three-layer unit cells in the validation dataset. As shown, the unit cells generated over different epochs are compared to the expected target unit cells. These plots can be assessed at the end of the training process and used to choose the best generator model based on the quality of generated unit cells. The figures show that the generated three-layer unit cells structures get more similar to the original unit cells as the number of training epochs increases. There are significant differences between the generated unit cells and the original ones after the initial epochs (e.g., after 5 or 25 epochs). In other words, the generator model is initially unable to mislead the PatchGAN model used as the discriminator for the cGAN model. However, by increasing the number of epochs to 50, the U-Net model generates unit cells that are similar to the target unit cells. Generating high-quality unit cells by the generator model deceives the discriminator to make mistake in distinguishing the fake generated unit cells from the real ones (i.e., by returning lower loss values for updating the generator model). Therefore, the generated unit cells begin to look realistic and get more similar to the original unit cells. For the rest of the training process, the generated unit cells' quality appears to improve or at least remain good. More importantly, after epoch 95, it can be seen that the cGAN discriminator is almost fooled by the generator and the discriminator model can not distinguish the original unit cells from the fake unit cells generated by the generator model. The figures show the same pattern for all three layers of the unit cells and also the rest of the unit cells in the training and validation datasets. It should be mentioned that training the cGAN model for more epochs does not necessarily lead to a better quality model, high-quality images (or more plausible designs for unit cells in this study). However, the cGAN model is trained up to 150 epochs in the present study to find the optimal number of epochs. Because the generated unit cells at epoch 95 demonstrate satisfying consistency to the original unit cells, this study chooses the trained model up to this epoch as the final model and uses it to design metasurfaces unit cells. In addition, at this epoch, the cGAN model performs well on both training and validation dataset and there is no concern about overfitting. Hence, this study uses this model for inverse design of EM microwave metasurfaces based on the procedure explained in step 6 of \textbf{Figure 4} given the desired scattering parameters. A case study example is presented in the next section based on this model.

\begin{figure*}[t]
	\centering
	\includegraphics[height=5in]{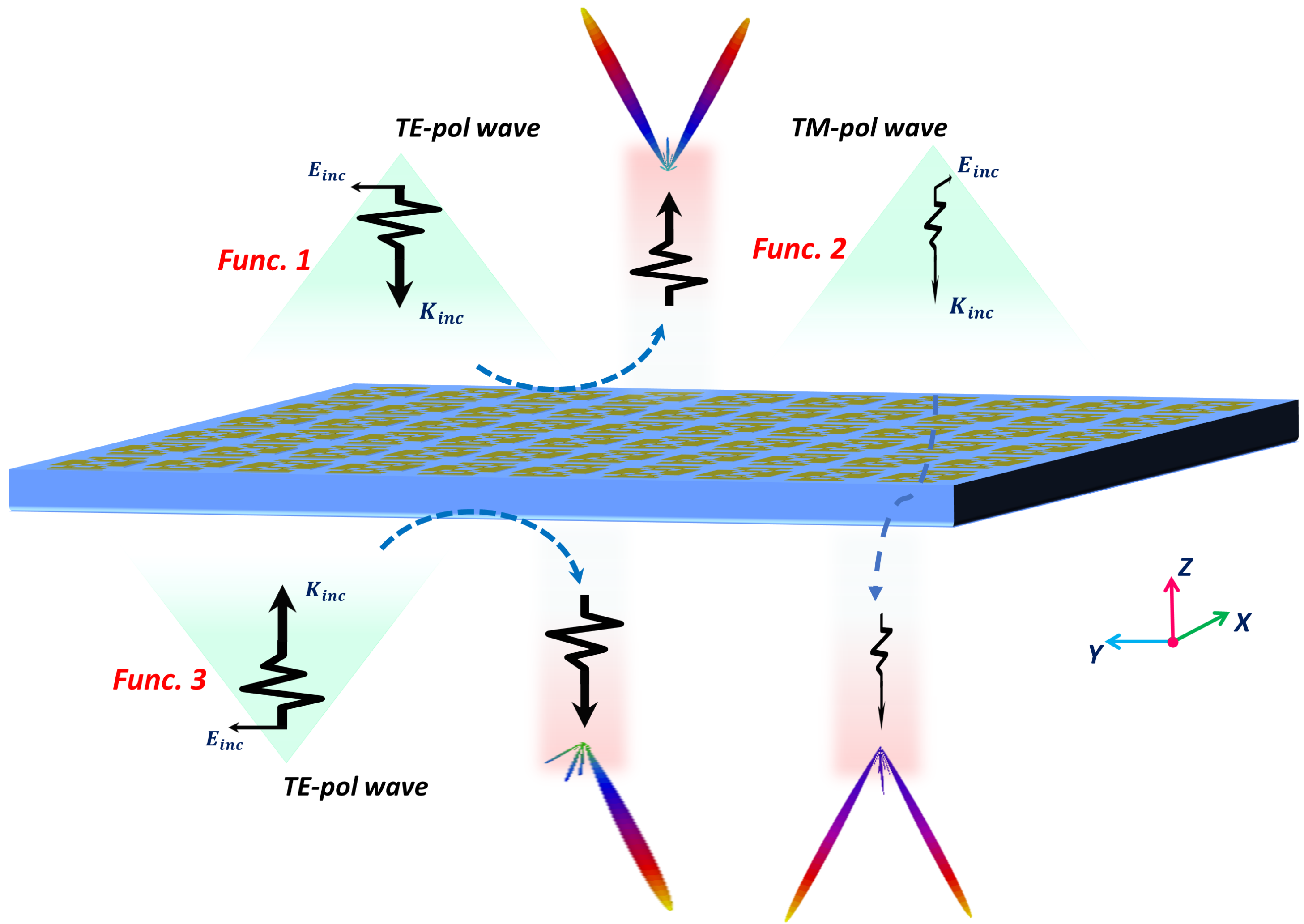}
	\caption{\label{fig:epsart} {The schematic illustration of the multi-functional metasurface designed using the cGAN model.}}
\end{figure*}
	\begin{figure*}[!t]
	\centering
	\includegraphics[height=5in]{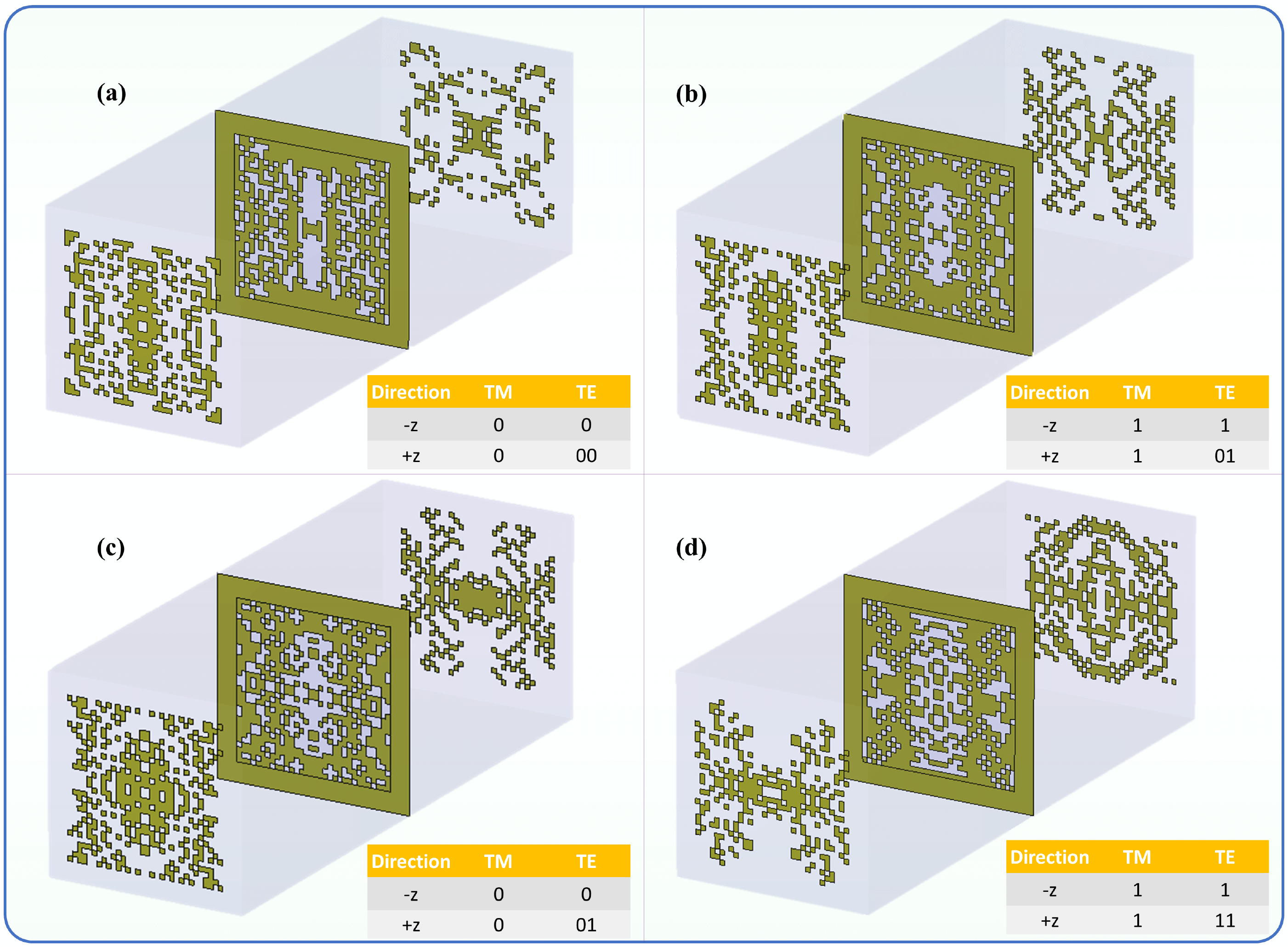}
	\caption{\label{fig:epsart} {The generated unit cells for realizing the multi-functional metasurface of Figure 8 with their coding states in TE and TM polarizations and  $ -z $ and $ +z $ directions.}}
\end{figure*}
\begin{figure*}[!t]
	\centering
	\subfigure[][]{%
		\label{fig:23}%
		\includegraphics[height=2.4in]{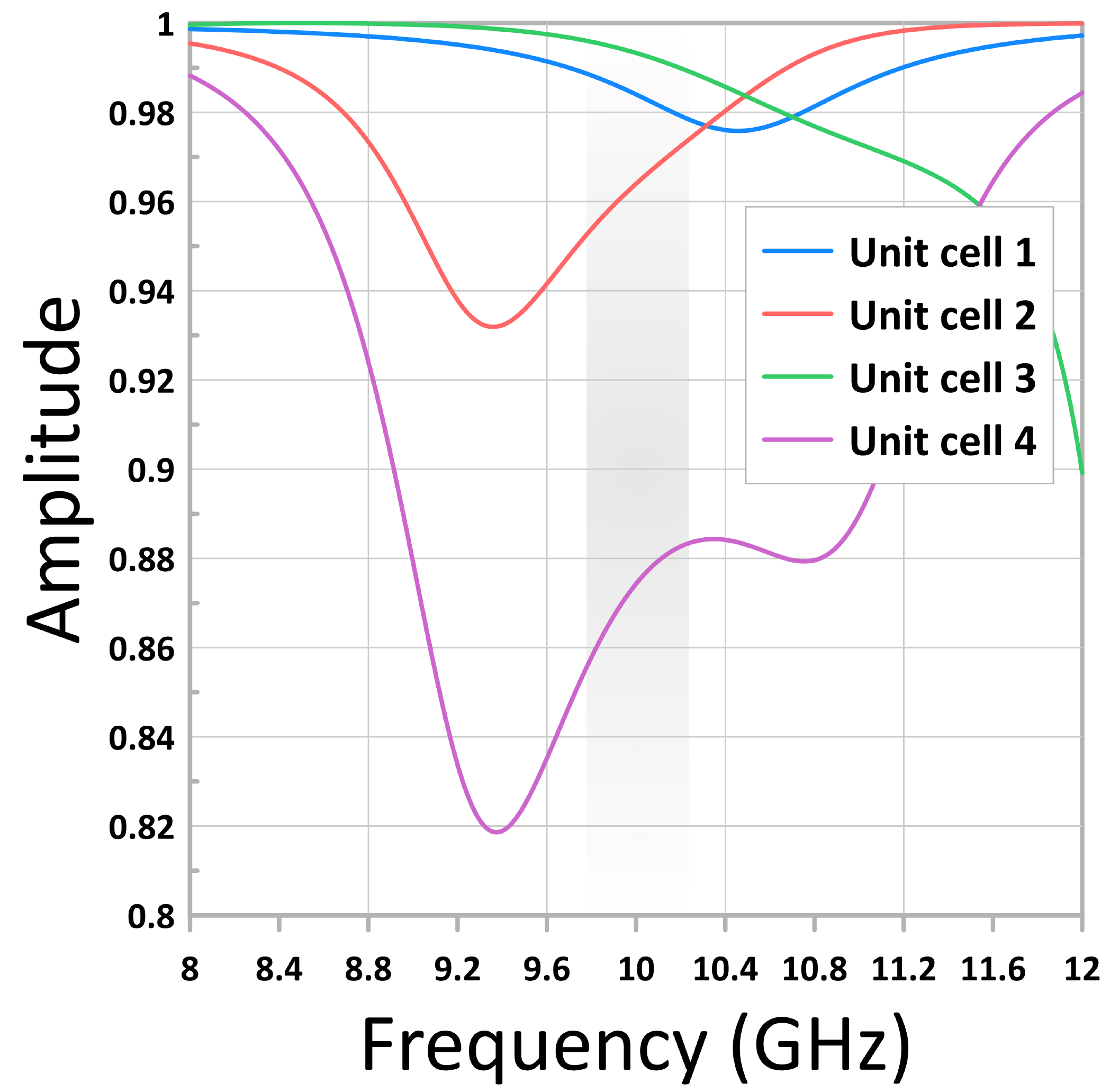}}%
	\subfigure[][]{%
		\label{fig:23}%
		\includegraphics[height=2.4in]{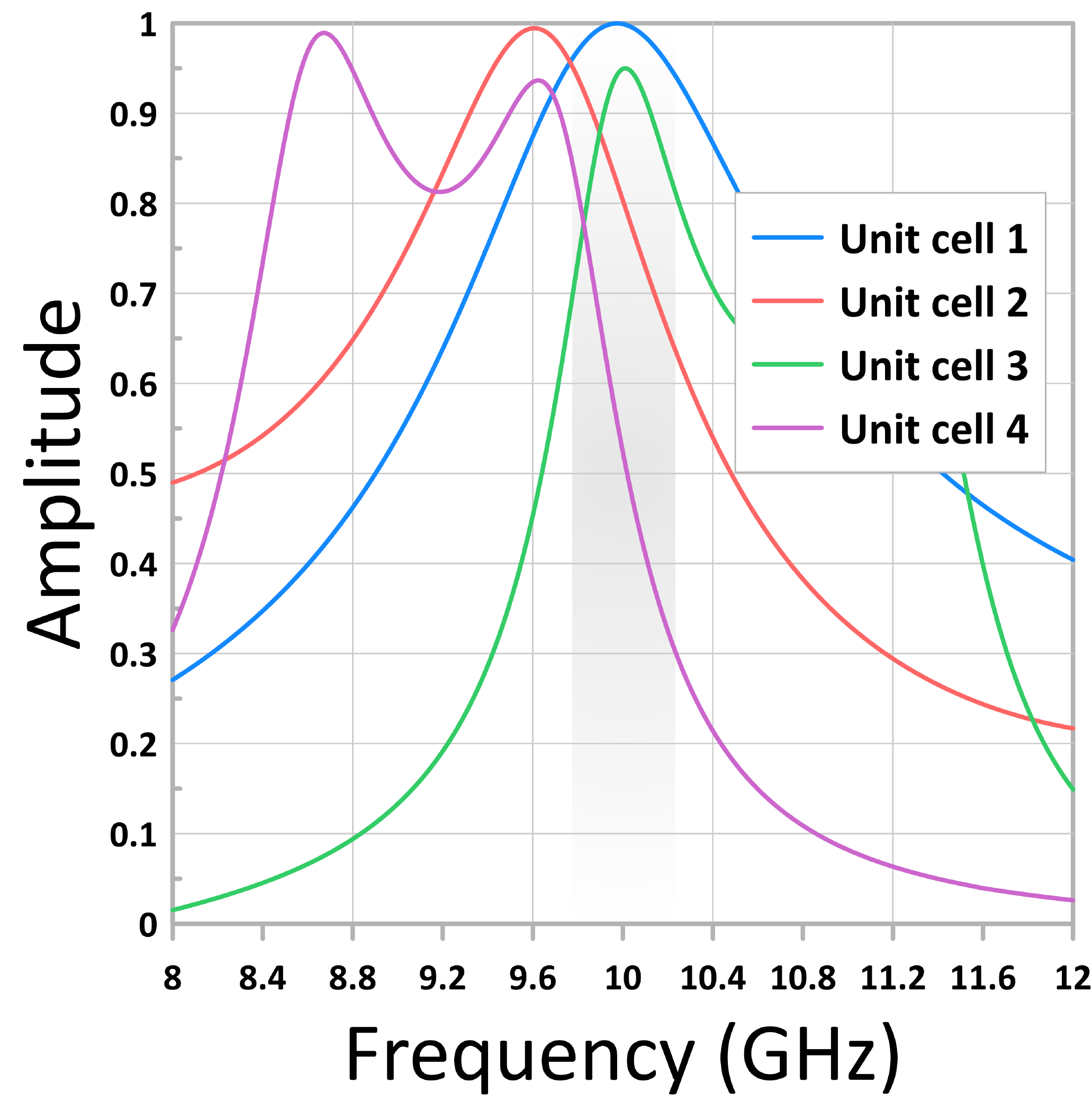}}%
	\subfigure[][]{%
		\label{fig:23}%
		\includegraphics[height=2.4in]{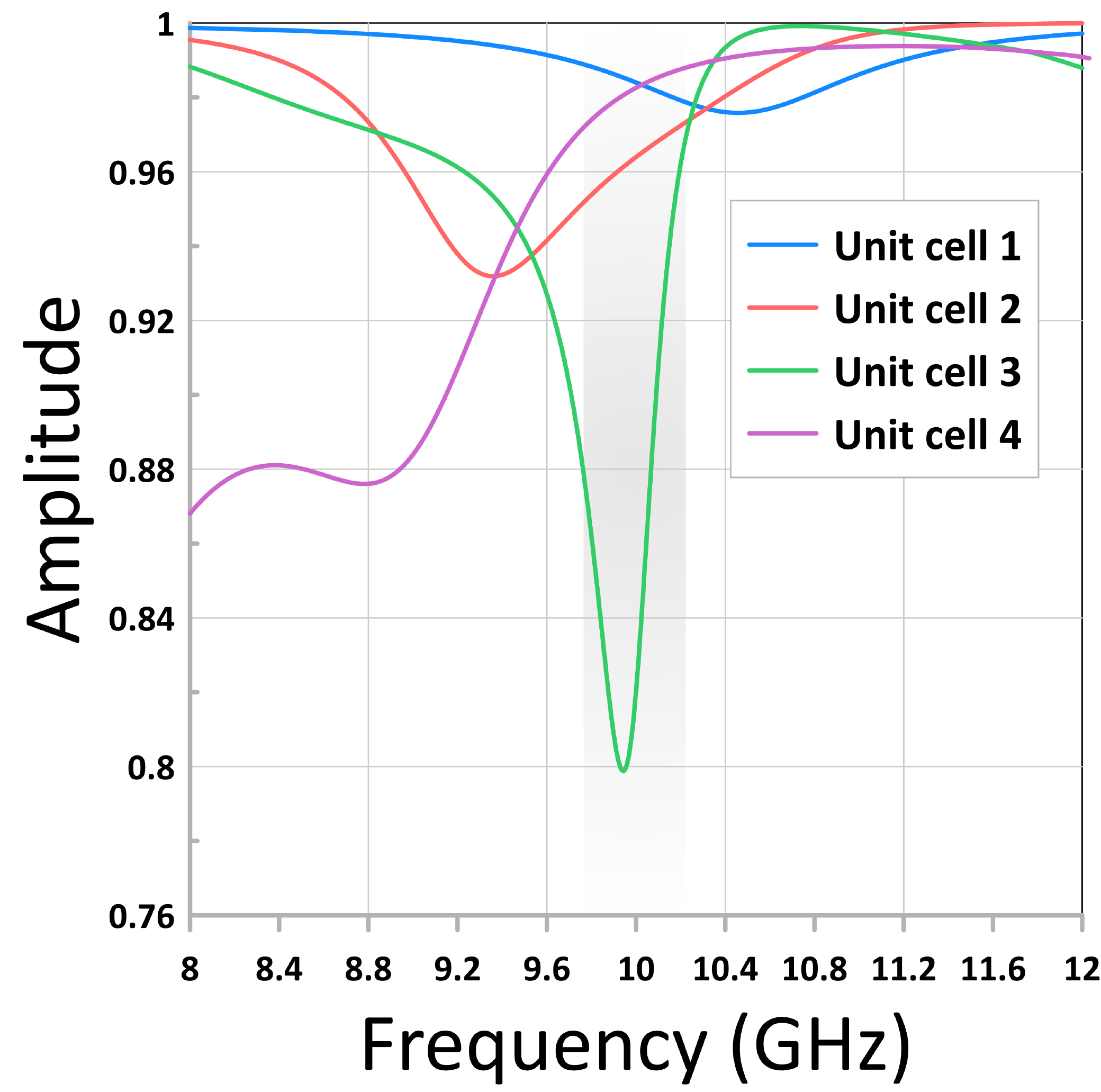}}%
	\qquad
	\subfigure[][]{%
		\label{fig:23}%
		\includegraphics[height=2.4in]{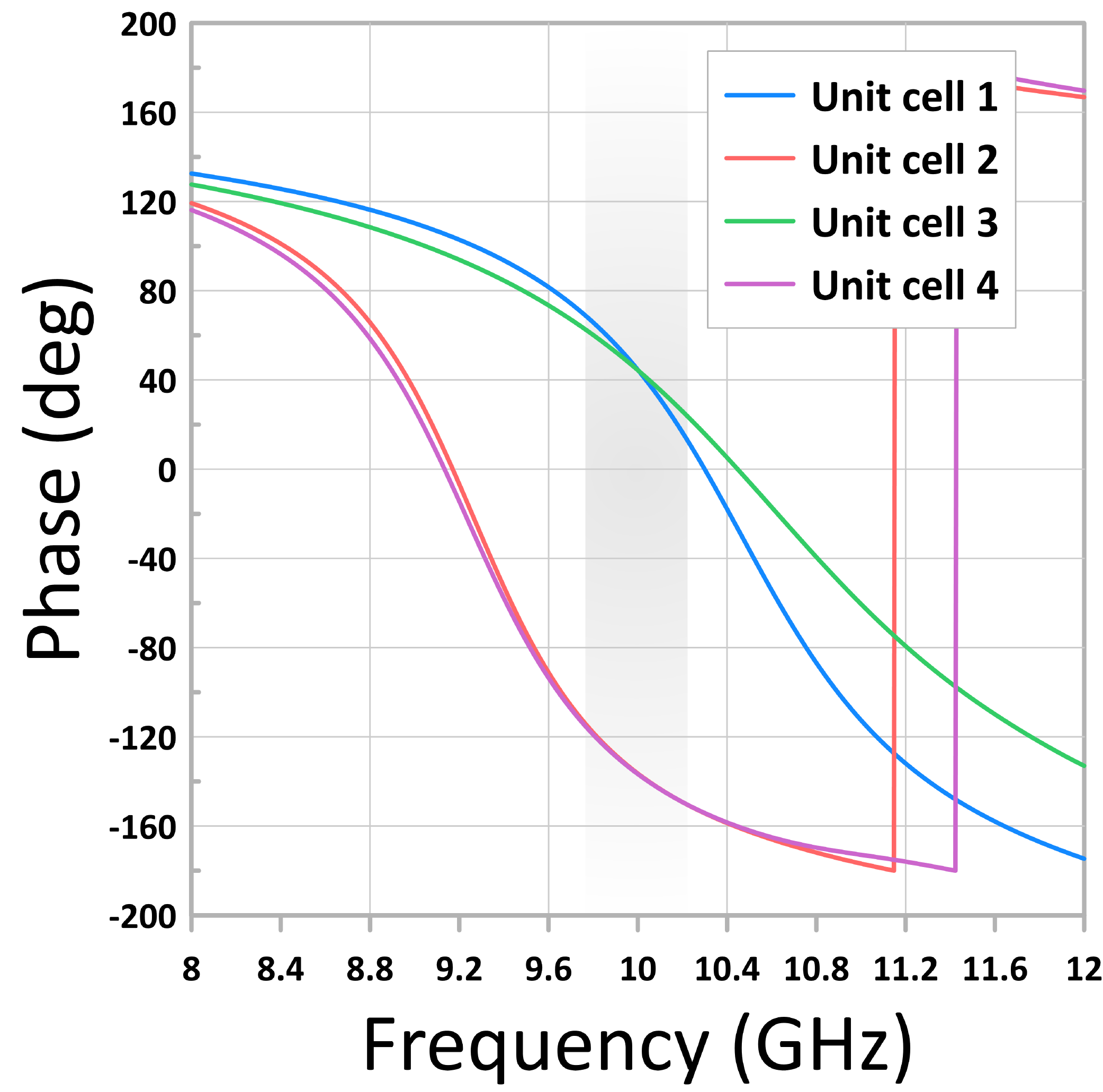}}%
	\subfigure[][]{%
		\label{fig:23}%
		\includegraphics[height=2.4in]{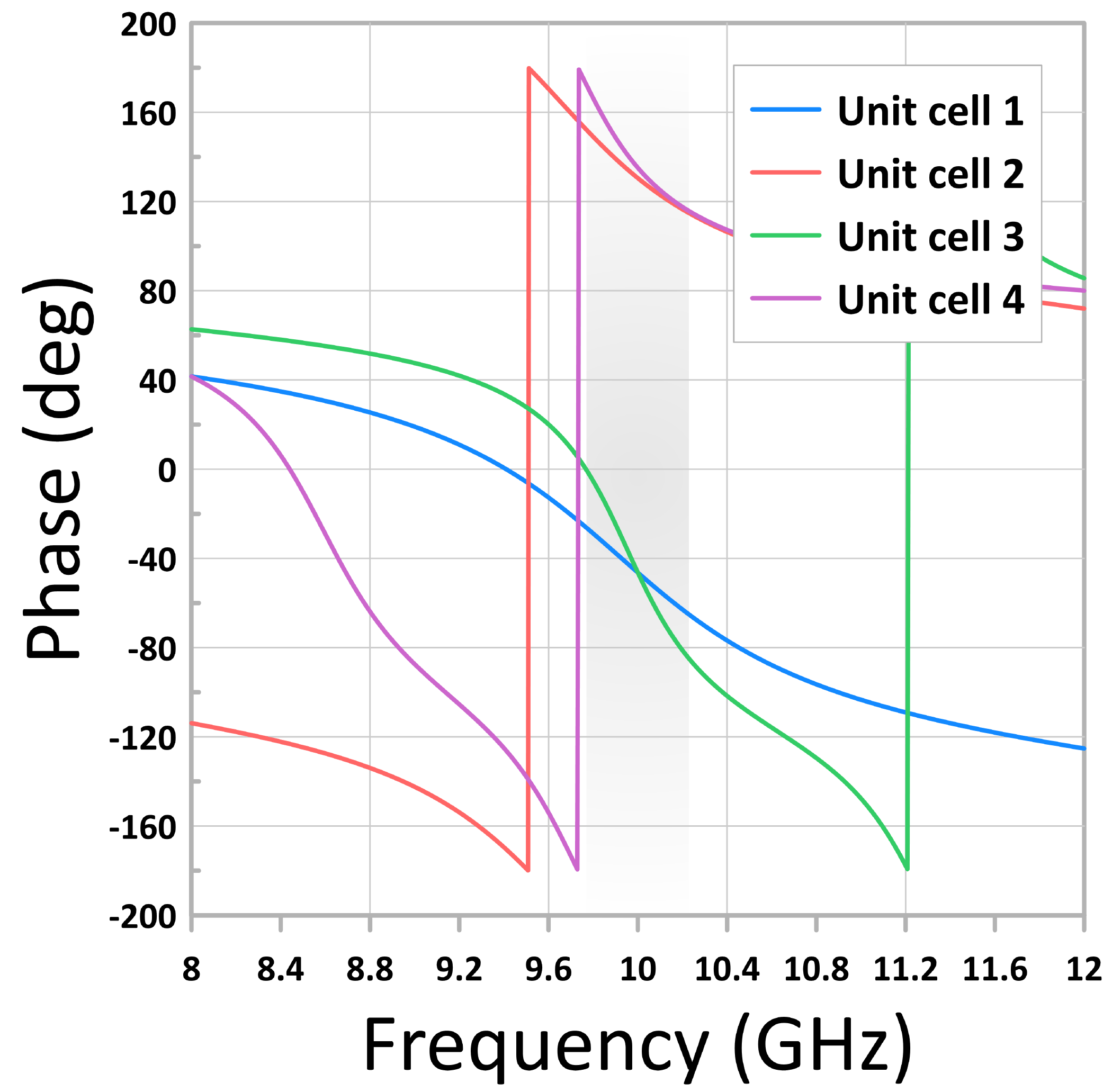}}%
	\subfigure[][]{%
		\label{fig:23}%
		\includegraphics[height=2.4in]{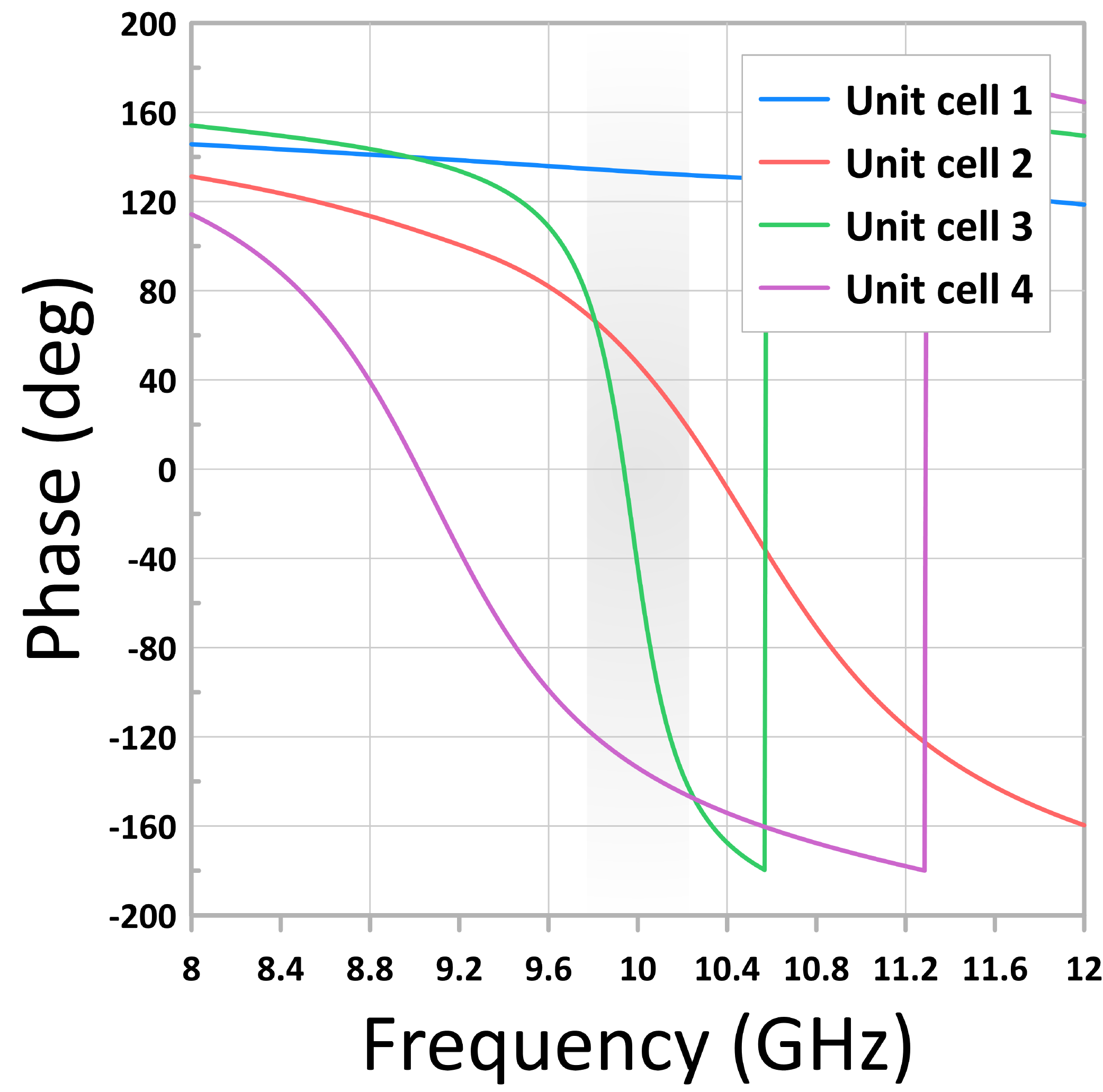}}%
	\caption{\label{fig:epsart} {The scattering parameters of the generated unit cells. (a), (c) The reflection coefficients amplitude in both $ -z $ and $ +z $ direction, respectively. (b) The transmission coefficient amplitude. (d), (e) The reflection coefficients phase in both $ -z $ and $ +z $ direction, respectively. (f) The transmission coefficient phase. }}
\end{figure*}

	\subsection{Metasurface Inverse Design Using cGAN Model}
	
		This part of the study presents a three-functional digital coding microwave metasurface, designed using the proposed cGAN model for validating its efficiency and performance in the inverse design of multi-functional metasurfaces. \textbf{Figure 8} exhibits the designed multi-functional metasurface that manipulates the incident EM waves in three independent ways in both transmitted and reflected modes at the same frequency band depending on the incoming wave polarization and direction. It is worth noting that it is possible to generate metasurfaces with multiple functions in different frequencies because the cGAN model generates unit cells in proportion to the three separate components of the scattering parameters in the whole frequency range of 8 GHz to 12 GHz. When the designed metasurface is normally illuminated by TM-polarized EM waves propagating along the $ -z $ or $ +z $ direction, it efficiently transmits the EM waves and displays tailored phases covering the whole $ 360^{\circ} $ range. While the metasurface reflects TE-polarized EM waves normally radiating from the $ -z $ or $ +z $ direction at the X frequency band (around 10 GHz). On the other hand, the reflection characteristics of the proposed metasurface depend on the incident wave direction so that the metasurface reveals independent functions in different propagation directions ($ -z $ and $ +z $ directions). 
		
		The proposed metasurface is incorporated different coding patterns from four types of unit cells that realize three functionalities (Func. 1 to Func. 3) at 10 GHz frequency. In the first and second functionalities (Func. 1 and Func. 2), the metasurface under the periodic coding sequence of 010101…/010101...,  reflects and transmits the normally incident TE- and TM-polarized EM waves propagating along $ -z $ direction to two symmetrically oriented directions, respectively. Whereas in the third functionality (Func. 3), the metasurface operates as a 2-bit coding surface when subjected to $ +z $ direction TE-polarized EM waves and under periodic coding sequence 0001101100011011…, the normally incident waves are reflected at oblique angles. 
		
		For the inverse design of such a multi-functional metasurface, the scattering parameters required to obtain the coding states at 10 GHz are initially converted to GAFs images using step 3 of \textbf{Figure 4}. After stacking (step 4 of \textbf{Figure 4}), the generated images are passed into the trained generator (step 6 of \textbf{Figure 4}). Then, the generator designs the corresponding unit cells structures shown in \textbf{Figure 9}. Each unit cell describes three coding states against TE-polarized EM waves propagating along the $ -z $ direction, TE-polarized EM waves propagating along the $ +z $ direction, and TM-polarized EM waves. For example, \textbf{Figure 9a} acts as “0” binary code for TE- and TM-polarized EM waves from $ -z $ direction and realizes a “00” binary code against TE-polarized EM waves from $ +z $ direction. Other required coding states for the desired applications are performed by the other three unit cells of \textbf{Figure 9b-d}. 
		
		For further verification, the generated unit cells are simulated using the CST Microwave Studio to calculate the reflection coefficients $ S_{11} $ and $ S_{22} $, as well as the transmission coefficient $ S_{21} $. At the frequency of the interest, 10 GHz, the scattering parameters amplitude, shown in \textbf{Figure 10a-c}, indicate that the reflections and transmission of the four designed unit cells subjected to EM waves are greater than 0.8. Moreover, \textbf{Figure 10d,e} presents when TE- and TM-polarized EM waves from $ -z $ direction are incident to generated unit cells, there is a $ 180^{\circ} $ phase difference between the reflection, as well as the transmission of two unit cells, 1 and 3, and other two unit cells, 2 and 4, respectively. Whereas according to \textbf{Figure 10f}, the rate of phase change among the unit cells exposed to TE-polarized waves from $ +z $ direction is $ 90^{\circ} $. Consequently, deploying the generated unit cells using the cGAN model in the specific configurations can produce a three-functional microwave passive metasurface for wave manipulation in the whole space (see \textbf{Figure 8}). The suggested cGAN model for the inverse design of multi-functional microwave metasurfaces performs well in numerical simulations. In fact, in a range of application scenarios, the cGAN-based technique provides an efficient method for multi-functional metasurface inverse design.
\section{Conclusion}
	A step-by-step framework is presented based on the cGANs for inverse design of multi-functional microwave metasurfaces to mitigate the computational costs of the conventional inverse design methods. The proposed methodology leverages a technique brought by the recent computer vision developments, known as GAFs, to initially encode the multi-objective scattering parameters of the three-layer unit cells as images to use them as the input for the cGAN model. In addition, the matrix representation of the unit cells structures is considered the target for the cGAN model. The cGAN model is comprised of two distinct convolutional neural networks competing against each other, including a conditional generator model for designing the unit cells given the desired scattering parameters and a discriminator model to determine whether the generated unit cells are plausible designs for the given scattering parameters. A robust study dataset including 54,000 three-layer unit cells with various configurations (i.e., various MSBPPs and MVSBAs) and their EM responses are implemented to train and validate the cGAN model. 
	
	The fully-trained cGAN model demonstrates its efficiency and capability in designing multi-functional metasurfaces via the design of a metasurface with three independent functions at a fixed frequency. The results of numerical EM case study simulations confirm that the proposed cGAN model can be adopted as an efficient tool for designing three-functional metasurfaces. Generally, the results suggest that the cGAN model is an efficient method to reduce the computational costs associated with the inverse design of multi-functional microwave metasurfaces. Although the results of this study are based on a specific dataset for metasurfaces with a particular shape, they are promising, and the same framework can be re-applied for designing multi-functional microwave metasurfaces with different shapes (e.g., circles, rectangles, “H”s, and quasi-free forms). Additionally, owing to the simple structure, the proposed methodology can be extended to develop multi-functional metadevices for many diversified application demands at microwave, terahertz, and optical regimes.
	


	
	
	%
	\bibliographystyle{MSP}
	\bibliography{template}

\begin{thebibliography}{10}
\providecommand{\url}[1]{\texttt{#1}}
\providecommand{\urlprefix}{URL }

\bibitem{chen2016review}
H.-T. Chen, A.~J. Taylor, N.~Yu,
\newblock \emph{Reports on progress in physics} \textbf{2016}, \emph{79}, 7
  076401.

\bibitem{li2018metasurfaces}
A.~Li, S.~Singh, D.~Sievenpiper,
\newblock \emph{Nanophotonics} \textbf{2018}, \emph{7}, 6 989.

\bibitem{chen2011mantle}
P.-Y. Chen, A.~Alu,
\newblock \emph{physical review B} \textbf{2011}, \emph{84}, 20 205110.

\bibitem{sounas2015unidirectional}
D.~L. Sounas, R.~Fleury, A.~Al{\`u},
\newblock \emph{Physical Review Applied} \textbf{2015}, \emph{4}, 1 014005.

\bibitem{zhao2011manipulating}
Y.~Zhao, A.~Al{\`u},
\newblock \emph{Physical Review B} \textbf{2011}, \emph{84}, 20 205428.

\bibitem{kiani2020spatial}
M.~Kiani, A.~Momeni, M.~Tayarani, C.~Ding,
\newblock \emph{Optics express} \textbf{2020}, \emph{28}, 23 35128.

\bibitem{monticone2013full}
F.~Monticone, N.~M. Estakhri, A.~Alu,
\newblock \emph{Physical review letters} \textbf{2013}, \emph{110}, 20 203903.

\bibitem{yu2011light}
N.~Yu, P.~Genevet, M.~A. Kats, F.~Aieta, J.-P. Tetienne, F.~Capasso,
  Z.~Gaburro,
\newblock \emph{science} \textbf{2011}, \emph{334}, 6054 333.

\bibitem{zheng2015metasurface}
G.~Zheng, H.~M{\"u}hlenbernd, M.~Kenney, G.~Li, T.~Zentgraf, S.~Zhang,
\newblock \emph{Nature nanotechnology} \textbf{2015}, \emph{10}, 4 308.

\bibitem{li2017electromagnetic}
L.~Li, T.~J. Cui, W.~Ji, S.~Liu, J.~Ding, X.~Wan, Y.~B. Li, M.~Jiang, C.-W.
  Qiu, S.~Zhang,
\newblock \emph{Nature communications} \textbf{2017}, \emph{8}, 1 197.

\bibitem{pors2015analog}
A.~Pors, M.~G. Nielsen, S.~I. Bozhevolnyi,
\newblock \emph{Nano letters} \textbf{2015}, \emph{15}, 1 791.

\bibitem{babaee2021parallel}
A.~Babaee, A.~Momeni, A.~Abdolali, R.~Fleury,
\newblock \emph{Physical Review Applied} \textbf{2021}, \emph{15}, 4 044015.

\bibitem{liu2017experimental}
X.~Liu, K.~Fan, I.~V. Shadrivov, W.~J. Padilla,
\newblock \emph{Optics express} \textbf{2017}, \emph{25}, 1 191.

\bibitem{kiani2020self}
M.~Kiani, M.~Tayarani, A.~Momeni, H.~Rajabalipanah, A.~Abdolali,
\newblock \emph{Optics Express} \textbf{2020}, \emph{28}, 4 5410.

\bibitem{aieta2012aberration}
F.~Aieta, P.~Genevet, M.~A. Kats, N.~Yu, R.~Blanchard, Z.~Gaburro, F.~Capasso,
\newblock \emph{Nano letters} \textbf{2012}, \emph{12}, 9 4932.

\bibitem{khorasaninejad2017metalenses}
M.~Khorasaninejad, F.~Capasso,
\newblock \emph{Science} \textbf{2017}, \emph{358}, 6367.

\bibitem{saenz2009coupling}
E.~Saenz, I.~Ederra, R.~Gonzalo, S.~Pivnenko, O.~Breinbjerg, P.~de~Maagt,
\newblock \emph{IEEE Transactions on Antennas and Propagation} \textbf{2009},
  \emph{57}, 2 383.

\bibitem{quevedo2015ultrawideband}
O.~Quevedo-Teruel, M.~Ebrahimpouri, M.~N.~M. Kehn,
\newblock \emph{IEEE Antennas and Wireless Propagation Letters} \textbf{2015},
  \emph{15} 484.

\bibitem{sievenpiper1999high}
D.~Sievenpiper, L.~Zhang, R.~F. Broas, N.~G. Alexopolous, E.~Yablonovitch,
\newblock \emph{IEEE Transactions on Microwave Theory and techniques}
  \textbf{1999}, \emph{47}, 11 2059.

\bibitem{krizhevsky2012imagenet}
A.~Krizhevsky, I.~Sutskever, G.~E. Hinton,
\newblock \emph{Advances in neural information processing systems}
  \textbf{2012}, \emph{25} 1097.

\bibitem{voulodimos2018deep}
A.~Voulodimos, N.~Doulamis, A.~Doulamis, E.~Protopapadakis,
\newblock \emph{Computational intelligence and neuroscience} \textbf{2018},
  \emph{2018}.

\bibitem{nadkarni2011natural}
P.~M. Nadkarni, L.~Ohno-Machado, W.~W. Chapman,
\newblock \emph{Journal of the American Medical Informatics Association}
  \textbf{2011}, \emph{18}, 5 544.

\bibitem{deng2013new}
L.~Deng, G.~Hinton, B.~Kingsbury,
\newblock In \emph{2013 IEEE international conference on acoustics, speech and
  signal processing}. IEEE, \textbf{2013} 8599--8603.

\bibitem{biamonte2017quantum}
J.~Biamonte, P.~Wittek, N.~Pancotti, P.~Rebentrost, N.~Wiebe, S.~Lloyd,
\newblock \emph{Nature} \textbf{2017}, \emph{549}, 7671 195.

\bibitem{krizhevsky2012advances}
A.~Krizhevsky, I.~Sutskever, G.~E. Hinton, F.~Pereira, C.~Burges, L.~Bottou,
  K.~Weinberger \textbf{2012}.

\bibitem{Kiani2019}
J.~Kiani, C.~Camp, S.~Pezeshk,
\newblock \emph{Computers and Structures} \textbf{2019}, \emph{218} 108.

\bibitem{campbell2020explosion}
S.~D. Campbell, R.~P. Jenkins, P.~J. O'Connor, D.~Werner,
\newblock \emph{IEEE Antennas and Propagation Magazine} \textbf{2020},
  \emph{63}, 3 16.

\bibitem{raccuglia2016machine}
P.~Raccuglia, K.~C. Elbert, P.~D. Adler, C.~Falk, M.~B. Wenny, A.~Mollo,
  M.~Zeller, S.~A. Friedler, J.~Schrier, A.~J. Norquist,
\newblock \emph{Nature} \textbf{2016}, \emph{533}, 7601 73.

\bibitem{Kiani2020}
J.~Kiani, C.~Camp, S.~Pezeshk, N.~Khoshnevis,
\newblock \emph{Computers and Structures} \textbf{2020}, \emph{241} 106355.

\bibitem{Ma2018}
W.~Ma, F.~Cheng, Y.~Liu,
\newblock \emph{ACS Nano} \textbf{2018}, \emph{12}, 6 6326.

\bibitem{zhou2021inverse}
M.~Zhou, D.~Liu, S.~W. Belling, H.~Cheng, M.~A. Kats, S.~Fan, M.~L. Povinelli,
  Z.~Yu,
\newblock \emph{ACS Photonics} \textbf{2021}, \emph{8}, 8 2265.

\bibitem{yeung2021global}
C.~Yeung, R.~Tsai, B.~Pham, B.~King, Y.~Kawagoe, D.~Ho, J.~Liang, M.~W. Knight,
  A.~P. Raman,
\newblock \emph{Advanced Optical Materials} \textbf{2021}, \emph{9}, 20
  2100548.

\bibitem{Li2019}
L.~Li, H.~Ruan, C.~Liu, Y.~Li, Y.~Shuang, A.~Al{\`{u}}, C.~W. Qiu, T.~J. Cui,
\newblock \emph{Nature Communications} \textbf{2019}, \emph{10}, 1.

\bibitem{nadell2019deep}
C.~C. Nadell, B.~Huang, J.~M. Malof, W.~J. Padilla,
\newblock \emph{Optics express} \textbf{2019}, \emph{27}, 20 27523.

\bibitem{liu2021intelligent}
C.~Liu, W.~M. Yu, Q.~Ma, L.~Li, T.~J. Cui,
\newblock \emph{Photonics Research} \textbf{2021}, \emph{9}, 4 B159.

\bibitem{an2021multifunctional}
S.~An, B.~Zheng, H.~Tang, M.~Y. Shalaginov, L.~Zhou, H.~Li, M.~Kang, K.~A.
  Richardson, T.~Gu, J.~Hu, et~al.,
\newblock \emph{Advanced Optical Materials} \textbf{2021}, \emph{9}, 5 2001433.

\bibitem{Qiu2019}
T.~Qiu, X.~Shi, J.~Wang, Y.~Li, S.~Qu, Q.~Cheng, T.~Cui, S.~Sui,
\newblock \emph{Advanced Science} \textbf{2019}, \emph{6}, 12.

\bibitem{Zhang2019}
Q.~Zhang, C.~Liu, X.~Wan, L.~Zhang, S.~Liu, Y.~Yang, T.~J. Cui,
\newblock \emph{Advanced Theory and Simulations} \textbf{2019}, \emph{2}, 2 1.

\bibitem{hodge2019rf}
J.~A. Hodge, K.~V. Mishra, A.~I. Zaghloul,
\newblock In \emph{2019 IEEE International Symposium on Phased Array System \&
  Technology (PAST)}. IEEE, \textbf{2019} 1--6.

\bibitem{naseri2021generative}
P.~Naseri, S.~V. Hum,
\newblock \emph{IEEE Transactions on Antennas and Propagation} \textbf{2021}.

\bibitem{mohammadjafari2021designing}
S.~Mohammadjafari, O.~Ozyegen, M.~Cevik, E.~Kavurmacioglu, J.~Ethier, A.~Basar,
\newblock \emph{Neural Computing and Applications} \textbf{2021}, 1--15.

\bibitem{shi2020metasurface}
X.~Shi, T.~Qiu, J.~Wang, X.~Zhao, S.~Qu,
\newblock \emph{Journal of Physics D: Applied Physics} \textbf{2020},
  \emph{53}, 27 275105.

\bibitem{zhu2021phase}
R.~Zhu, T.~Qiu, J.~Wang, S.~Sui, C.~Hao, T.~Liu, Y.~Li, M.~Feng, A.~Zhang,
  C.-W. Qiu, et~al.,
\newblock \emph{Nature communications} \textbf{2021}, \emph{12}, 1 1.

\bibitem{cui2014coding}
T.~J. Cui, M.~Q. Qi, X.~Wan, J.~Zhao, Q.~Cheng,
\newblock \emph{Light: Science \& Applications} \textbf{2014}, \emph{3}, 10
  e218.

\bibitem{momeni2018information}
A.~Momeni, K.~Rouhi, H.~Rajabalipanah, A.~Abdolali,
\newblock \emph{Scientific reports} \textbf{2018}, \emph{8}, 1 1.

\bibitem{liu2017concepts}
S.~Liu, T.~J. Cui,
\newblock \emph{Advanced Optical Materials} \textbf{2017}, \emph{5}, 22
  1700624.

\bibitem{rouhi2021multi}
K.~Rouhi, S.~E. Hosseininejad, S.~Abadal, M.~Khalily, R.~Tafazolli,
\newblock \emph{Journal of Lightwave Technology} \textbf{2021}, \emph{39}, 21
  6893.

\bibitem{tang2020wireless}
W.~Tang, M.~Z. Chen, X.~Chen, J.~Y. Dai, Y.~Han, M.~Di~Renzo, Y.~Zeng, S.~Jin,
  Q.~Cheng, T.~J. Cui,
\newblock \emph{IEEE Transactions on Wireless Communications} \textbf{2020},
  \emph{20}, 1 421.

\bibitem{li2017nonlinear}
A.~Li, Z.~Luo, H.~Wakatsuchi, S.~Kim, D.~F. Sievenpiper,
\newblock \emph{IEEE Access} \textbf{2017}, \emph{5} 27439.

\bibitem{momeni2022switchable}
A.~Momeni, K.~Rouhi, R.~Fleury,
\newblock \emph{Carbon} \textbf{2022}, \emph{186} 599.

\bibitem{CST}
CST,
\newblock Cst microwave studio advanced topics,
\newblock Technical report, CST-Computer Simulation Technology, \textbf{2002}.

\bibitem{goodfellow2014generative}
I.~Goodfellow, J.~Pouget-Abadie, M.~Mirza, B.~Xu, D.~Warde-Farley, S.~Ozair,
  A.~Courville, Y.~Bengio,
\newblock \emph{Advances in neural information processing systems}
  \textbf{2014}, \emph{27}.

\bibitem{mirza2014conditional}
M.~Mirza, S.~Osindero,
\newblock \emph{arXiv preprint arXiv:1411.1784} \textbf{2014}.

\bibitem{isola2017image}
P.~Isola, J.-Y. Zhu, T.~Zhou, A.~A. Efros,
\newblock In \emph{Proceedings of the IEEE conference on computer vision and
  pattern recognition}. \textbf{2017} 1125--1134.

\bibitem{zhu2017unpaired}
J.-Y. Zhu, T.~Park, P.~Isola, A.~A. Efros,
\newblock In \emph{Proceedings of the IEEE international conference on computer
  vision}. \textbf{2017} 2223--2232.

\bibitem{wu2018pool}
D.~Wu,
\newblock \emph{IEEE transactions on neural networks and learning systems}
  \textbf{2018}, \emph{30}, 5 1348.

\bibitem{wang2015imaging}
Z.~Wang, T.~Oates,
\newblock In \emph{Twenty-Fourth International Joint Conference on Artificial
  Intelligence}. \textbf{2015} .

\bibitem{ronneberger2015u}
O.~Ronneberger, P.~Fischer, T.~Brox,
\newblock In \emph{International Conference on Medical image computing and
  computer-assisted intervention}. Springer, \textbf{2015} 234--241.

\bibitem{szegedy2017inception}
C.~Szegedy, S.~Ioffe, V.~Vanhoucke, A.~A. Alemi,
\newblock In \emph{Thirty-first AAAI conference on artificial intelligence}.
  \textbf{2017} .

\bibitem{srivastava2014dropout}
N.~Srivastava, G.~Hinton, A.~Krizhevsky, I.~Sutskever, R.~Salakhutdinov,
\newblock \emph{The journal of machine learning research} \textbf{2014},
  \emph{15}, 1 1929.

\bibitem{bengio2012practical}
Y.~Bengio,
\newblock In \emph{Neural networks: Tricks of the trade}, 437--478. Springer,
  \textbf{2012}.

\bibitem{kingma2014adam}
D.~P. Kingma, J.~Ba,
\newblock \emph{arXiv preprint arXiv:1412.6980} \textbf{2014}.

\bibitem{salimans2016improved}
T.~Salimans, I.~Goodfellow, W.~Zaremba, V.~Cheung, A.~Radford, X.~Chen,
\newblock \emph{Advances in neural information processing systems}
  \textbf{2016}, \emph{29} 2234.

\end{thebibliography}
\end{document}